\documentclass[12pt]{article}

\usepackage{natbib}
\usepackage[nottoc,notlof,notlot]{tocbibind} 

\bibpunct{(}{)}{;}{a}{}{;}
\usepackage{breakcites}
\usepackage[breaklinks=true]{hyperref}
\hypersetup{
    colorlinks=true,
    linkcolor=blue,
    citecolor=blue,
    urlcolor=blue
}

\usepackage{graphicx}
\graphicspath{ {images/} }

\usepackage{xcolor}

\usepackage{amsmath}
\usepackage{bbm}
\usepackage{amsfonts}%
\usepackage{amssymb}%
\usepackage{amsthm}

\usepackage{graphicx} 
\usepackage{framed}
\usepackage{caption}
\usepackage{subcaption}

\usepackage{booktabs} 
\usepackage{bigstrut} 
\usepackage{array} 
\usepackage{paralist} 
\usepackage{verbatim} 

\usepackage{lipsum}
\usepackage{setspace}

\usepackage[singlelinecheck=false]{caption}
\usepackage{tabularx}

\usepackage{multirow}
\usepackage{lscape}
\usepackage[linesnumbered,ruled,vlined]{algorithm2e}

\title{\setstretch{1} A Modified Randomization Test for the \\ Level of Clustering}

 \date{\parbox{\linewidth}{\centering%
  \today\endgraf}} 

\theoremstyle{definition}

\newtheorem{theorem}{Theorem}

\newtheorem{assumption}{Assumption}

\newtheorem{remark}{Remark}

\newcommand{\blind}{0}

\begin{document}

\def\spacingset#1{\renewcommand{\baselinestretch}%
{#1}\small\normalsize} \spacingset{1}


\if0\blind
{
  \title{\bf A Modified Randomization Test for the Level of Clustering}
  \author{Yong Cai\thanks{I am grateful to Ivan Canay for extensive guidance on this project. I thank Eric Auerbach, Eduardo Campillo Betancourt, Grant Goehring and Joel Horowitz for helpful comments.}\hspace{.2cm}\\
    Department of Economics, Northwestern University\\
    \url{yongcai2023@u.northwestern.edu}
   }
  \maketitle
} \fi

\if1\blind
{
  \bigskip
  \bigskip
  \bigskip
  \begin{center}
    {\LARGE\bf A Modified Randomization Test for the \\[5pt]  Level of Clustering}
\end{center}
  \medskip
} \fi

\bigskip
\begin{abstract}
Suppose a researcher observes individuals within a county within a state. Given concerns about correlation across individuals, it is common to group observations into clusters and conduct inference treating observations across clusters as roughly independent. However, a researcher that has chosen to cluster at the county level may be unsure of their decision, given knowledge that observations are independent across states. This paper proposes a modified randomization test as a robustness check for the chosen level of clustering in a linear regression setting. Existing tests require either the number of states or number of counties to be large. Our method is designed for settings with few states and few counties. While the method is conservative, it has competitive power in settings that may be relevant to empirical work.
\end{abstract}

\noindent%
\vfill

\newpage
\spacingset{1.45} 


\section{Introduction}

Consider the following regression:
\begin{equation*}
	Y_i = X_i'\beta + U_i \quad , \quad E[X_{i}U_{i}] = 0~.
\end{equation*}
where a researcher wants to perform inference on $\beta$. If the researcher is concerned about correlation between $U_{i}$ and $U_{i'}$, it is frequently helpful to group observations into independent clusters. These independent clusters can then be used to construct cluster-robust covariance estimators (CCE) as in \cite{lz1986}, or for approximate randomization tests as in \cite{crs2017} and \cite{ccks2021}. 


However, these procedures require the assignment of units to clusters be known ex ante. In practice, researchers often have some freedom in choosing the level at which to cluster their standard errors. For example, those working with the American Community Survey (ACS) can cluster their data either at the individual, county or state level. Alternatively, those working with firm data from COMPUSTAT have the option to cluster firms at either the 4-digit, 3-digit or 2-digit Standard Industrial Classification (SIC) level. 

Clustering at the correct level is important for valid inference. A large body of simulation evidence shows that ignoring cluster dependence -- in other words, clustering at too fine a level -- leads to type I errors that exceed the nominal error by as much as 10 times (\citealt{bdm2004}; \citealt{cgm2008}). On the other hand, clustering at excessively coarse levels can also lead to problems. For one, coarse clusters tend to be few in number. It is well-known that confidence intervals based on the cluster-robust standard errors tend to under-cover when the number of clusters is small (see \citealt{mhe2008} for instance), leading to poor size control. In the absence of under-coverage issues, unnecessarily coarse levels of clustering can also lead to tests with poor power since the researcher assumes less information than they actually have. \citealt{aaiw17} demonstrate via simulations, in a many-cluster setting, that CCEs based on coarse clusters can be too large. They also provide theoretical results in this vein, though they do so in the context of their ``design-based" asymptotics that differ from those traditionally used to analyze clustered standard errors. Nonetheless, the problems with tests based on excessively coarse-clustering arise even with few clusters -- the setting of interest for our paper. We present a simple simulation to demonstrate these issues in Appendix \ref{appendix--inferecefewclusters}. 

Given the above considerations, a researcher may choose to cluster at a fine level (e.g. individual or county) even when a coarse level of clustering (e.g. state), which is known to be valid, is also available. Nonetheless, they may be unsure if the fine level is appropriate. That is, whether observations across the fine clusters are approximately independent. 

To help researchers assess the validity of their chosen clusters, we propose a modified randomization test that can be used as a robustness check for a given clustering specification. Our test requires large (fine) sub-clusters, but is justified under asymptotics that take the number of (coarse) clusters and (fine) sub-clusters as fixed. Inference is difficult in this setting because scores are not independent across sub-clusters even asymptotically, as we will explain in Section \ref{section--teststatcritval}. Randomization tests, which typically require some type of asymptotic independence, thus cannot be directly applied. We get around this problem by searching for worst-case values of the unobserved parameters to guard against over-rejection. We describe a simple method to search for this value, so that the computational complexity of the test is of the same order as the number of sub-clusters. This is reasonable since our test is targeted towards applications with few sub-clusters. Our test has no power against negative correlation. However, ignoring negative correlation leads to variance estimators that are too large, and is thus less of an issue if the researcher is concerned about size control when performing inference on $\beta$. 

To our knowledge, there are two other tests for the level of clustering. \cite{mnw2020} proposes a test based on having large number of coarse clusters, relying on the wild bootstrap to improve finite sample performance. Meanwhile, \cite{im2016} proposes a test for the case when there are many sub-clusters. Our test, which takes the number of clusters and sub-clusters to be fixed, handles a more challenging situation, though this comes at the cost of being conservative, especially in settings with homogeneous clusters. However, as our simulations in Section \ref{section--montecarlo} show, it has competitive power given heterogeneous clusters -- a setting that could be relevant for empirical work. Indeed, our test detects correlation in the clusters chosen by \cite{gllqsx2019}, demonstrating its potential usefulness in applied work (see Section \ref{section--application}). Finally, we note that the test of \cite{im2016} also has no power against negative correlation, although that of \cite{mnw2020} does not share this limitation.

\cite{aaiw17} takes a different approach to this issue. They argue for a ``design-based" perspective on clustering, requiring researchers to determine ex ante the uncertainty that they face in either sampling or treatment assignment. For example, if the researcher believes that in their specific context, treatment assignment occurs at the sub-cluster level, then sub-clusters should be used for computing standard errors, regardless of whether or not residuals are correlated across the sub-clusters. While insightful, this approach requires researchers to answer an alternative question on which there is equally little theoretically guidance. We therefore develop our method under the ``model-based" framework, in which the researcher has in mind some data-generating process that entails dependent clusters. 

The remainder of this paper is organized as follows. Section \ref{section--proposedtest} describes our proposed test. Section \ref{section--montecarlo} presents Monte Carlo simulations. Section \ref{section--application} demonstrates an application to \cite{gllqsx2019}. Section \ref{section--conclusion} concludes. Proofs are collected in Appendix \ref{appendix--proofs}.

\section{The Proposed Test}\label{section--proposedtest}

\subsection{Model and Assumptions}
In the following, we assume that the researcher has conducted inference on $\beta \in \mathbf{R}$, and seeks a robustness check for the level of clustering used for said inference. As will become clear in Section \ref{section--implementation}, using a scalar $\beta$ yields computational advantages, though the test can be feasibly computed for moderate dimensions of $\beta$. For this reason and for ease of exposition we limit our discussion to the scalar case. 

Consider the linear regression:
\begin{equation}\label{equation--model}
Y_{i} = X_{i}\beta + W_{i}'\gamma +  U_{i}, \quad E[X_{i}U_{i}] = 0, \, E[W_{i}U_{i}] = 0~,
\end{equation}
where $\beta \in \mathbf{R}$ is the parameter of interest and $\gamma \in \mathbf{R}^d$ is a nuisance parameter. Suppose there are $r$ clusters, indexed by $k \in \mathcal{K}$. Within each cluster $k$, there are $q_k$ sub-clusters, indexed by $j \in \mathcal{J}_k$. Within each sub-cluster $j$, there are $n_j$ individuals indexed by $i \in \mathcal{I}_j$. Let $\mathcal{J} = \bigcup_{k \in \mathcal{K}} \mathcal{J}_k$ and $\mathcal{I} = \bigcup_{j \in \mathcal{J}} \mathcal{I}_j$. Further, let $n = \sum_{j \in \mathcal{J}} n_j$ and $q = \sum_{k \in \mathcal{K}} q_k = |\mathcal{J}|$. We also write $i \in \mathcal{I}_k$ when $i \in \mathcal{I}_j$ and $j \in \mathcal{J}_k$. In the following, we suppress dependence on $j$ and $k$ whenever this does not cause confusion.

\begin{assumption}\label{assumption--Xprojection} 
Suppose that for every cluster $j$, there exists a vector $\Pi_j$, with a consistent estimator $\hat{\Pi}_j$, such that for all $i \in \mathcal{I}_j$:
\begin{equation}\label{equation--pi}
	X_i = W_i'\Pi_j + \varepsilon_i, \quad E[W_i\varepsilon_i] = 0~.
\end{equation}
\end{assumption}

Suppose that within a sub-cluster, $W$ has full rank. Then $\hat{\Pi}_j$ can be chosen as the sub-cluster level OLS estimator of $X$ on $W$. Otherwise, we can just drop variables until we obtain a linearly independent subset $\tilde{W}$. The entries of $\hat{\Pi}_j$ corresponding to the dropped variables can then be set to $0$ while the remaining entries are chosen to be the corresponding coefficients from the sub-cluster level regression of $X$ on $\tilde{W}$. Alternatively, if the researcher is willing to assume that $\Pi_j$ is identical across clusters, $\hat{\Pi}_j$ can also be obtained from the full sample regression of $X$ on $W$. Now define
\begin{equation}\label{definition--z}
	Z_i = (X_i - W_i\Pi_j){U}_i \quad \mbox{ and } \quad \hat{Z}_i = (X_i - W_i\hat{\Pi}_j)\hat{U}_i~,
\end{equation}
where $\hat{U}_i$ is the full-sample OLS residual using equation (\ref{equation--model}). Suppose we know that clusters are independent, so that $E[Z_{i}Z_{i'}] = 0$ when $i \in \mathcal{I}_k, i' \in \mathcal{I}_{k'}$ and $k \neq k'$. Under this assumption, we test the null hypothesis that sub-clusters are uncorrelated:
\begin{equation}\label{equation--nullhypothesis}
H_0: E[Z_{i}Z_{i'}] = 0 \mbox{ for all } i \in \mathcal{I}_{j}, i' \in \mathcal{I}_{j'}, j \neq j'
\end{equation}
against the alternative hypothesis that there exists sub-clusters within at least one cluster that exhibit correlation:
\begin{equation*}
H_A: E[Z_{i}Z_{i'}] \neq 0 \mbox{ for some } i \in \mathcal{I}_j, i' \in \mathcal{I}_{j'} \mbox{ such that } j, j' \in \mathcal{J}_k, j \neq j'~.
\end{equation*}


Note that changing the choice of $X_i$ and $W_i$ corresponds to testing different null hypotheses and could lead to differing outcomes. If a researcher wants to test the level of clustering used for inference on $\beta$, $X_i$ should be projected onto $W_i$. Similarly, if inference was conducted on $\gamma$, then $W_i$ should take the place of $X_i$ in equation (\ref{definition--z}). 

\begin{remark}
	A researcher interested in inference on $\beta$ only has to test the residualized hypothesis of equation (\ref{equation--nullhypothesis}). This is because to the first order, the asymptotic distribution of
	\begin{equation*}
		\sqrt{n}\left(\hat{\beta} - \beta\right) \quad \text{and} \quad \sqrt{n_j}\left(\hat{\beta}_j - \beta\right)
	\end{equation*}
	depends only on 
	\begin{equation*}
		\frac{1}{\sqrt{n}} \sum_{i \in \mathcal{I}} Z_i \quad \text{and} \quad \frac{1}{\sqrt{n_j}} \sum_{i \in \mathcal{I}_j} Z_i~.
	\end{equation*}
	respectively. Hence, if the $Z_i$'s exhibit no correlation across clusters, then conducting inference using the sub-clusters is appropriate. We flesh out this argument in Appendix \ref{section--residualizednull}. The fact that tests for different coefficients require different adjustments for clustering is unsurprising. A similar phenomenon arises in methods employing degrees of freedom correction for inference with a small number of clusters. Here, each slope parameter in a regression may require a test with different degrees of freedom (see \cite{ik2016} and \cite{bm2002}). 
\end{remark}


\begin{remark}
	As with \cite{im2016} and \cite{mnw2020}, we require the researcher to specify independent clusters which nest the potentially correlated sub-clusters. While this is not always feasible, researchers seeking to test the level of clustering typically have a few choices available to them. Coarser clusters are also frequently considered to be more believably independent than the finer ones. It is natural to apply our tests in these instances.  
\end{remark}

We further assume the following:
\begin{assumption}\label{assumption--clt}
	Suppose $q$ and $r$ are fixed, but $n_j \to \infty$ for all $j \in \mathcal{J}$. Let ${Z}_{i}$ be defined as in equation (\ref{definition--z}). Suppose there exists $\Omega \in \mathbf{R}^{q \times q}$ such that the $q$-vector $S_n \overset{d}{\to} S$, where
	\begin{equation} \label{equation--Sn_S}
	S_n := \begin{pmatrix}
	\frac{1}{\sqrt{n_1}} \sum_{i \in \mathcal{I}_1} {Z}_{i} \\
	\vdots \\
	\frac{1}{\sqrt{n_{q}}} \sum_{i \in \mathcal{I}_{q}} {Z}_{i}
	\end{pmatrix}
	\quad \mbox{ and } \quad S := N \left( \mathbf{0}, \Omega \right).
	\end{equation}
	Further, let $\hat{\beta}$ and $\hat{\gamma}$ be the (joint) respective OLS estimators of $\beta$ and $\gamma$, as defined in equation (\ref{equation--model}) and $\hat{\Pi}_j$ be the estimator of $\Pi_j$ as defined in equation (\ref{equation--pi}). Suppose: 
	\begin{equation*}
	\hat{\beta} \overset{p}{\to} \beta, \quad  \hat{\gamma} \overset{p}{\to} \gamma, \quad \sqrt{n_j}\left(\hat{\Pi}_j  - \Pi_j\right) = O_p(1) \quad \mbox{ for all } j \in \mathcal{J}~.
	\end{equation*}
\end{assumption}

In other words, we assume that the errors are weakly correlated within each sub-cluster $j$. Imposing weak dependence within a (sub-)cluster is not an uncommon assumption (see for instance the discussion in \cite{crs2017} and \cite{bch2011}). We note that under $H_0$, $\Omega$ is a diagonal matrix. On the other hand, under the alternative, it has a block diagonal structure due to correlation between sub-clusters. 

\begin{remark}
	Our assumption that $S_n \to S$ does not implicitly assume that (sub-)clusters have similar sizes. Intuitively, this is because our randomization test assigns ``equal weight" to each sub-cluster: each sub-cluster is normalized by its own $n_j$, and the sign of each $S_{n,j}$ contribute equally to the sign mismatch within its parent cluster. As such, heterogeneous sub-cluster sizes pose no issue for our test. Nonetheless, the quality of the asymptotic approximation is determined by $\min_{j \in [q]} n_j$, so the smallest cluster has to be large. We expand on this point in Appendix \ref{section--impliedheterogeneity} and explain how the restricted heterogeneity assumptions that are required for inference with clustered data are not needed in our case. 
\end{remark}

\begin{remark}
	Without further assumptions on $Z_i$, our test requires large sub-clusters. This rules out testing the null of no clustering where there is only one observation in each sub-cluster. However, the test is valid for the null of no clustering if we are willing to assume that each $Z_i$ is symmetrically distributed around $0$. Such assumptions can be found in the econometrics literature. For example, \cite{df2008} use it to justify a wild-bootstrapped based $F$-test for the linear regression model. Nonetheless, we consider this assumption to be highly restrictive and hence justify our test via large sub-cluster asymptotics.
\end{remark}

\subsection{Test Statistic and Critical Value}\label{section--teststatcritval}

In this subsection, we define the test statistic and explain the need to search over the worst case critical value. Before doing so, we first consider the infeasible test in which the true parameters -- $\beta$, $\gamma$ and $\Pi$ as defined in equations (\ref{equation--model}) and (\ref{equation--pi}) -- are observed. Readers who are only interested in the details of implementation can skip to the end of Section \ref{section--implementation}. 

\subsubsection{Infeasible Test}\label{subsubsection--infeasibletest}

Suppose we know $\beta$, $\gamma$ and $\Pi$. Given $Y_i$ and $X_i$, we can back out $U_i$ and construct the vector $S_n^*$, whose $j^\text{th}$ entry is 
\begin{equation}\label{equation--Sstar}
	S^*_{n,j} =  \frac{1}{\sqrt{n_j}} \sum_{i \in \mathcal{I}_j} Z_i = \frac{1}{\sqrt{n_j}} \sum_{i \in \mathcal{I}_j} \left(X_i - W_i'{\Pi}_j\right) {U}_{i}~.
\end{equation}
Given $S^*_{n}$, we can then define the infeasible test statistic:
\begin{equation}\label{eqn--infeasibleteststat}
T(S^*_n) = \frac{1}{r}\sum_{k \in \mathcal{K}} \left\lvert \sum_{j \in \mathcal{J}_k} \left(\mathbf{1} ({S}^*_{n,j} \geq 0) - \mathbf{1} ({S}^*_{n,j}< 0) \right) \right\rvert~.
\end{equation}
The inner sum is the net number of positive ${S}^*_{n,j}$ within each cluster $k$. Intuitively, if the sub-clusters are independent, the net number of positive $S^*_{n,j}$ should be close to $0$. Conversely, if they are positively correlated, this number will be large in absolute value, since many sub-clusters will have $S_{n,j}^*$ of the same sign. On the other hand, if they are negatively correlated, this number will be more concentrated around $0$ than in the independent case. As will become clear below, our test interprets large absolute values of $T(S_n^*)$ as violation of the null. For this reason, we it will not have power against negative correlation.

\begin{remark}
	There are two advantages to having a test statistic that depends only on the sign of the $S_{n,j}^*$'s. Firstly, large and small realizations of $S_{n,j}^*$ contribute the same amount to $T(S_n^*)$. As such, the performance of our test is not affected even if sub-clusters have wildly differing variances, a source of heterogeneity that may be important in applied work. We demonstrate this robustness property via simulations in Section \ref{subsection--clusthet}. Secondly, the feasible version of this test requires searching over the worst case values of the test statistic. As will become clear in Section \ref{section--implementation}, this search is simplified by our choice of test statistic. 
\end{remark}

Next, denote by $\mathbf{G}$ the set of sign changes. $\mathbf{G}$ can be identified with the set of $g \in \{-1, 1\}^{q}$ so that 
\begin{equation*}
g{S}^*_n = 	\begin{pmatrix}
g_1 \cdot {S}^*_{n,1} \\
\vdots \\
g_{q} \cdot {S}^*_{n,q}
\end{pmatrix}~.
\end{equation*}
Now let $p^*(S_n^*)$ be the proportion of $T\left(g{S}^*_n\right)$ that are no smaller than $T\left({S}^*_n\right)$:
\begin{equation}\label{equation--pstar}
	p(S_n^*) = \frac{1}{\lvert\mathbf{G}\rvert} \sum_{g \in \mathbf{G}} \mathbf{1}\left\{ T\left(g{S}^*_n\right) \geq T\left({S}^*_n\right) \right\}~.
\end{equation}
The test rejects the null hypothesis when $p(S_n^*)$ is small -- that is, when $T\left({S}^*_n\right)$ is extreme relative to $T\left(g{S}^*_n\right)$:
\begin{align} \label{test_infeasible}
\phi^*_n = \begin{cases}
1 & \mbox{ if } p(S_n^*) \leq \alpha \\
0 & \mbox{ otherwise.}
\end{cases}
\end{align}

The intuition for the randomization test is as follows. Since $S^*_{n,j}$ involves only units within the same sub-cluster, under the null hypothesis, $S_n^*$ converges to a mean-zero normal distribution with independent components. Independence, together with symmetry of normal random variables about their means, implies that for any $g \in \mathbf{G}$, $gS^*_n$, has the same distribution as $S_n^*$. Hence, the randomization distribution $\{T(gS_n^*)\}_{g\in \mathbf{G}}$ is in fact the distribution of $T(S_n^*)$ conditional on the values of $|S_n^*|$, where $|\cdot|$ is applied component-wise. Rejecting the null hypothesis when we observe values of $T(S_n^*)$ that are extreme relative to $\{T(gS_n^*)\}_{g\in \mathbf{G}}$ therefore leads to a test with the correct size. 

Note that the randomization test defined above is non-randomized. Randomization tests can also employ a randomized rejection rule for the situation when
\begin{equation*}
	\frac{1}{\lvert\mathbf{G}\rvert} \sum_{g \in \mathbf{G}} \mathbf{1}\left\{ T\left(g{S}^*_n\right) > T\left({S}^*_n\right) \right\} < \alpha \quad \mbox{ but } \quad  \frac{1}{\lvert\mathbf{G}\rvert} \sum_{g \in \mathbf{G}} \mathbf{1}\left\{ T\left(g{S}^*_n\right) \geq T\left(g{S}^*_n\right) \right\} > \alpha~.
\end{equation*}
Using a randomized rejection rule, we have that provided the necessary symmetry properties hold in finite sample, the randomization test will have size equal to $\alpha$ exactly. The test defined in equation (\ref{test_infeasible}) is conservative since it never rejects when the above situation occurs. However, we present the deterministic version since the test that we propose is based on it.

\subsubsection{Na\"ive Test}\label{subsubsection--naivetest}

Tests based on $S_n^*$ are infeasible since $\beta$, $\gamma$ and the $\Pi_j$'s are unknown. Suppose we simply replaced $Z_i$ with $\hat{Z}_i$ and performed the randomization test with the estimated scores. It turns out that this procedure is incorrect. To see this, let $\tilde{S}_n$ be $S_n^*$ but with $\hat{Z}_i$ replacing $Z_i$. Then we can write each component of $\tilde{S}_n$ as:
\begin{align}
	\tilde{S}_{n,j} & = \frac{1}{\sqrt{n_j}} \sum_{i \in \mathcal{I}_j} \hat{Z}_i = \frac{1}{\sqrt{n_j}} \sum_{i \in \mathcal{I}_j} \left(X_i - W_i'\hat{\Pi}_j\right) \hat{U}_{i} \\ 
	& \approx \underbrace{\frac{1}{\sqrt{n_j}} \sum_{i \in \mathcal{I}_j} \left(X_i - W_i'{\Pi}_j\right) {U}_{i}}_\text{\normalsize $S^*_{n,j}$} - \underbrace{\left(\hat{\beta} - \beta \right) \frac{1}{\sqrt{n_j}} \sum_{i \in \mathcal{I}_j} \left(X_i - W_i'\hat{\Pi}_j\right)^2}_\text{\normalsize $=: A_j$}~. \label{equation--intuition}
\end{align}
In the above equation, $S_{n,j}^*$ is the part that is informative about cluster structure. However, each component now has an additional nuisance term $A_j$ that does not go away under asymptotics that take the number of sub-clusters to be fixed. Because $\hat{\beta} - \beta$ is common across the $A_j$'s, it induces correlation across $\tilde{S}_{n,j}$ even when the $S_{n,j}^*$'s are independent, leading potentially to over-rejection. Addressing this complication which does not arise in frameworks taking $q \to \infty$ results in the conservativeness of our test. 



\subsubsection{Feasible Test}\label{subsubsection--feasibletest}

If we knew $\hat{\beta} - \beta$, we could back out ${S}^*_{n,j}$ for the randomization test using equation (\ref{equation--intuition}). Since that is not possible, we propose to search over values of $\hat{\beta} - \beta$ to ensure that the test controls size when the unobserved term takes on extreme values. 

For a given $\lambda \in \mathbf{R}$, let $\hat{S}_n({\lambda})$ be $q\times 1$ vector whose $j^\text{th}$ entry is the following term:
	\begin{equation*} \label{equation--Sn_tilde}
	\hat{S}_{n,j}({\lambda}) := 
	\frac{1}{\sqrt{n_j}} \sum_{i \in \mathcal{I}_j} \left(X_i - W_i'\hat{\Pi}_j\right) \hat{U}_{i} + 
	\lambda \frac{1}{\sqrt{n_j}} \sum_{i \in \mathcal{I}_j} \left(X_i - W_i'\hat{\Pi}_j\right)^2~.
	\end{equation*}
Note that $\hat{S}_{n,j}(\hat{\beta} - \beta) = S^*_{n,j} + o_p(1)$. Define:
\begin{equation*}
T(\hat{S}_n(\lambda)) = \frac{1}{r}\sum_{k \in \mathcal{K}} \left\lvert \sum_{j \in \mathcal{J}_k} \left(\mathbf{1} (\hat{S}_{n,j}(\lambda) \geq 0) - \mathbf{1} (\hat{S}_{n,j}(\lambda) < 0) \right)\right\rvert~.
\end{equation*}
For a given $\lambda$, this is just the test statistic in equation (\ref{eqn--infeasibleteststat}) but with $\hat{S}_{n,j}(\lambda)$ taking the place of $S_{n,j}^*$. As before, we denote by $\mathbf{G}$ the set of sign changes and write: 
\begin{equation*}
g\hat{S}_n(\lambda) = 	\begin{pmatrix}
g_1 \cdot \hat{S}_{n,1}(\lambda) \\
\vdots \\
g_{q} \cdot \hat{S}_{n,q}(\lambda)
\end{pmatrix}~.
\end{equation*}
Now let $p(\hat{S}_n(\lambda))$ be the proportion of $T\left(g\hat{S}_n(\lambda)\right)$ that takes on extreme values relative to $T\left(\hat{S}_n(\lambda)\right)$:
\begin{equation}\label{equation--plambda}
	p(\hat{S}_n(\lambda)) = \frac{1}{\lvert\mathbf{G}\rvert} \sum_{g \in \mathbf{G}} \mathbf{1}\left\{ T\left(g\hat{S}_n(\lambda)\right) > T\left(\hat{S}_n(\lambda)\right) \right\}~.
\end{equation}
We can then define the randomization test as:
\begin{align} \label{test_standard}
\phi_n = \begin{cases}
1 & \mbox{ if } \sup_{\lambda \in \mathbf{R}} p(\hat{S}_n(\lambda)) \leq \alpha \\
0 & \mbox{ otherwise.}
\end{cases}
\end{align}

We can then prove the following result:
\begin{theorem} \label{proposition--size}
	Under assumptions \ref{assumption--Xprojection} and \ref{assumption--clt}, $\underset{n \to \infty}{\lim\sup} \,\,\mathbb{E}[\phi_n] \leq \alpha$. 
\end{theorem}

The test is a two-stage process. In the first stage, it searches for the value of $\lambda$ that leads to the largest $p$-value. In the second stage, the test rejects if this worst-case $p$-value is still smaller than the desired level of significance $\alpha$. Since the worst-case $p$-value bounds the true $p$-value from above, the rejection rule based on the worst-case $p$-value must be conservative.

As the Monte Carlo simulations in Section \ref{section--montecarlo} shows, the test has size that could be much smaller than $\alpha$ under the null hypothesis. However, the same simulations also show that the test has reasonable power under the alternative hypothesis, particularly in settings where clusters are heterogeneous in their variances. The potential usefulness of our test is further seen in the empirical application (Section \ref{section--application}), where it detects dependence in the clusters chosen by \cite{gllqsx2019}. 


\begin{remark}
	The worst-case test has no power if $r = 1$ since $\lambda = \text{median}(\{\tilde{S}_{n,j}\})$ will set exactly half the signs of $\hat{S}_{n,j}(\lambda)$ to be positive and half to be negative, so that the signs are completely balanced. 
	However, this is no longer true with $r > 1$ since only a single value can be chosen to balance signs across multiple clusters. The implementation procedure provides further intuition for power in this test. See the next subsection. 
\end{remark}

\begin{remark}
	As with standard randomization tests, $|\mathbf{G}|$ may sometimes be too large so that computation of $p(\hat{S}_n(\lambda))$ becomes onerous. In these instances, it is possible to replace $p(\hat{S}_n(\lambda))$ with a stochastic approximation. Formally, let 
	$
		\hat{\mathbf{G}} = \left\{ g^1, ... ,g^B \right\}~,
	$
	where $g^1$ is the identity transformation and $g^2, ..., g^B$ are i.i.d. Uniform($\mathbf{G}$). Using $\hat{\mathbf{G}}$ instead of ${\mathbf{G}}$ in equation (\ref{equation--plambda}) does not affect validity of theorem \ref{proposition--size}. For implementation, we follow \cite{crs2017} in evaluating the $p(\hat{S}_n(\lambda))$ completely when $q \leq 10$ and approximating it with $B = 1000$ when $q > 10$. 
\end{remark}

\begin{remark}\label{remark--pretest}
	We advocate the use of our test as a robustness check, after a researcher has chosen a level of clustering for inference, in the same spirit that manipulation tests are routinely used in studies with regression discontinuity designs, or in tests for pre-trends in studies involving difference-in-differences. In particular, the original inference results should be presented with results of the current test, regardless of the outcome. Conceptually, this is different from using the test as a pre-test to select the level of clustering prior to inference. The distinction is important as pre-testing is known to induce uniformity issues, where inference in the second stage (on $\beta$) suffers from distortion due to mistakes in the pre-test (that happen with positive probability). These same concerns are articulated by \cite{im2016}, who argue that their test “merely provides empirical evidence on the plausibility of one particular clustering assumption”. We take exactly the same view of our test.
\end{remark}


\subsection{Implementation}\label{section--implementation}

In this subsection we describe an efficient way of searching for $\lambda \in \mathbf{R}$. This search is simplified by the fact that $p(\hat{S}_n(\lambda))$ depends only on the sign of $\hat{S}_{n,j}(\lambda)$'s. As such, to find $\sup_{\lambda \in \mathbf{R}}$, we only need to search over sign combinations of $\hat{S}_{n,j}$. When $\beta$ is scalar, the search can be completed in $O(q)$ time. This is reasonable since the test is designed for use when $q$ is small. 

Suppose for now that $\sum_{i \in \mathcal{I}_j} \left(X_i - W_i'\hat{\Pi}_j\right)^2 > 0$ for all $j \in \mathcal{J}$. Define: 
\begin{align*}
	R_j = \frac{\sum_{i \in \mathcal{I}_j} \left(X_i - W_i'\hat{\Pi}_j\right) \hat{U}_{i} }{ \sum_{i \in \mathcal{I}_j} \left(X_i - W_i'\hat{\Pi}_j\right)^2}~.
\end{align*}
Then, 
$
	\hat{S}_{n,j}({\lambda}) \geq 0 \Leftrightarrow R_j + \lambda \geq 0~.
$
Sort the values of $R_j$'s so that
$
	R^{(1)} \geq R^{(2)} \geq ... \geq R^{(q)}~.
$
We must have that $R^{(j)} +\lambda \geq 0 \Rightarrow R^{(j')} +\lambda \geq 0$ for all $j' \leq j$. Let $\hat{S}_{n, (1)}(\lambda), ..., \hat{S}_{n, (q)}(\lambda)$ denote the values of $\hat{S}_{n,j}(\lambda)$ corresponding to $R^{(1)}, ..., R^{(q)}$. Therefore, we only need to consider sequences of the form
\begin{equation*}
\hat{S}_{n, (1)} > 0, ..., \hat{S}_{n, (j)} > 0, \quad \hat{S}_{n, (j+1)} < 0, ..., \hat{S}_{n, (q)} < 0~,
\end{equation*}
for some cut-off $j$. Since the $p$-value, as defined in equation (\ref{equation--plambda}), depends only on the sign of $\hat{S}_n$, we can compute it using $\check{S}_n$ in the place of $\hat{S}_{n,j}(\lambda)$:
\begin{equation*}
\check{S}_{n, (1)} = ... = \check{S}_{n, (j)} = 1, \quad \check{S}_{n, (j+1)} = ... = \check{S}_{n, (q)} = -1~.
\end{equation*}

Here, we see that even when we are searching over the worst case $\lambda$, we are only allowed to choose the cut-off point at which the signs change. We can therefore complete the search with no more than $q$ randomization tests. Assuming that the time it takes for each test is $O(1)$, the procedure takes $O(q)$ time. The restriction that $\check{S}_{n, (j)} \geq \check{S}_{n, (j')}$ for all $j \leq j'$ also gives the test power. If all combinations of signs for the $S_{n,j}$'s were allowed, the test will always return a $p$-value of 1 and will have no power. 

Finally, suppose there are sub-clusters such that $\sum_{i \in \mathcal{I}_j} \left(X_i - W_i'\hat{\Pi}_j\right)^2 = 0$. We can repeat the above procedure excluding these sub-clusters. In the final step, we set $\check{S}_{n,j}$ corresponding to these clusters to 0. Hence, 

\begin{remark}
	We can further reduce computation time by the following. Let $$R^+_k =\min_{j' \in \mathcal{J}_k} \left\{R_{j'} \text{ greater than or equal to $>0.5$ of } \{R_j, j \in \mathcal{J}_k\} \right\}$$ be the ``upward-conservative" median. Also define the ``downward-conservative" median: $$R^-_k = \max_{j' \in \mathcal{J}_k} \left\{R_{j'} \text{ less than or equal to $>0.5$ of } \{R_j, j \in \mathcal{J}_k\} \right\}~.$$ 
	Now let $R^+ = \max_{k \in \mathcal{K}} R_k^+ \quad \mbox{and} \quad R^- = \min_{k \in \mathcal{K}} R_k^-~.$
	We only need to consider cutoffs below $R^+$. Setting the sign cutoff at the argmax of $R^+$ results in situation in which all clusters have at least half of their entries being $-1$. If we now set the extreme $S_{n,j}$'s to $-1$, this will increase the net number of $-1$'s in \emph{all} clusters. Since our test is based on sign imbalance within clusters, such sequences will lead to a strictly larger test statistic and smaller $p$-values than if they were set to 1. For the same reason, we only need to consider cutoffs above $R^-$. 
\end{remark}

We summarise the implementation procedure in Algorithm \ref{algorithm--WCR}.

\begin{algorithm}
\DontPrintSemicolon
  Perform full sample OLS to obtain residuals $\hat{U}_i$. Compute $\hat{\Pi}_j$ for each $j \in \mathcal{J}$.
  
  \For{$j \in [q]$}{
  	\lIf{$\sum_{i \in \mathcal{I}_j} \left(X_i - W_i'\hat{\Pi}_j\right)^2 > 0$}{compute:
  	$
  	  	R_j = \frac{\sum_{i \in \mathcal{I}_j} \left(X_i - W_i'\hat{\Pi}_j\right) \hat{U}_{i} }{ \sum_{i \in \mathcal{I}_j} \left(X_i - W_i'\hat{\Pi}_j\right)^2}
  	$}
  \lElse{set $R_j = 0$.} }
  
  Sort the values of $R_j$'s such that
$  	R^{(1)} \geq R^{(2)} \geq ... \geq R^{(q)}~. $
  
  \For{$j \in [q]$, $R_{(j)} \neq 0$, $R^- \leq R_{(j)} \leq R^+$}    
  { 
   		 Set $
         \check{S}_{n, (1)} = ... = \check{S}_{n, (j)} = 1, \quad \check{S}_{n, (j+1)} = ... = \check{S}_{n, (q)} = -1.
         $
         
         \lIf{$R_{(j)} = 0$}{replace $\check{S}_{n, (j)}$ with $0$.}
         
         Compute $p(\check{S}_n)$. This is as defined in equation (\ref{equation--plambda}), except with $\check{S}_n$ in place of $\hat{S}_{n,j}(\lambda)$. Save this value as $\hat{p}_j$. 
  }

  \lIf{$\max_{j \in [q], R_j \neq 0} \hat{p}_j \leq \alpha$}{return $1$. Reject the null hypothesis.} \lElse
    {
    	return $0$. Do not reject the null hypothesis. 
    }

\caption{Worst-Case Randomization Test}\label{algorithm--WCR}
\end{algorithm}

\subsection{Comparison with Existing Tests}\label{section--competingtests}

To our knowledge, two other tests have been proposed for the level of clustering. They take either the number of sub-clusters in each cluster to infinity or the number of clusters to infinity. We assume both to be fixed. For ease of exposition, we restrict our discussion of these tests to the univariate case. 

\cite{im2016} (IM hereafter) adopts an asymptotic framework that takes $q_k \to \infty$ for all $k \in \mathcal{K}$. Consider estimating a regression coefficient cluster-by-cluster. Let $\hat{\beta}_k$ denote coefficients estimated using only cluster $k$. The IM test is based on the asymptotic distribution of an estimator for the variance of $\frac{1}{r} \sum_{k = 1}^r\hat{\beta}_k$. Let this variance be denoted by $V$ and let $\hat{\Omega}^\text{CCE}_k$ be the cluster-robust variance estimator for $\hat{\beta}_k$, where the clustering is done at the sub-cluster level using $j \in \mathcal{J}_k$. Under the null hypothesis, $\hat{\Omega}^\text{CCE}_k$ consistently estimates the variance of each $\hat{\beta}_k$. 

Under either the null or the alternative, but maintaining the assumption that coarse clusters are independent, consider estimating $V$ by:
\begin{equation*}
	\hat{V} = \frac{1}{r-1} \sum_{k = 1}^r (\hat{\beta}_k - \bar{\beta})^2~, \qquad \bar{\beta} = \frac{1}{r} \sum_{r=1}^k \hat{\beta}_r~.
\end{equation*}
IM show that under the null, $\hat{V} \overset{d}{\to} V^W$, where $V^W = \frac{1}{r-1} \sum_{k = 1}^r ({W}_k - \bar{W})^2$ and 
\begin{equation*}
	W \sim N(0, \text{diag}({\Omega}^\text{CCE}_1, {\Omega}^\text{CCE}_2, ..., {\Omega}^\text{CCE}_r))~.
\end{equation*}
The IM test constructs a reference distribution $\hat{V}^W$ by drawing $W$ from $$N(0, \text{diag}(\hat{\Omega}^\text{CCE}_1, \hat{\Omega}^\text{CCE}_2, ..., \hat{\Omega}^\text{CCE}_r))$$ and seeing if $\hat{V}$ is larger than the $\left( 1-{\alpha}\right)^\text{th}$ quantile of $\hat{V}^W$. 

There are two limitations to the IM test that our test does not share. Firstly, they require the regression to be estimated cluster-by-cluster. This would be infeasible in, for example, differences-in-differences set ups where treatment varies at the cluster level. Secondly, since their asymptotics take $q_k \to \infty$, we expect the test to have poor properties when $q_k$ is small. Instead, our test is expected to have good properties even when $q_k$ is small as long as $n_j$ is large. These benefits come at a cost. We expect our test to perform worse if observations within sub-clusters are highly correlated, whereas the IM test allows unrestricted covariance within sub-clusters. Our test is also conservative under the null hypothesis. We note also that neither test has power against negative correlations. This is because both tests use test statistics that take on large value relative to their reference distributions only when there is positive correlation. 

\cite{mnw2020} (MNW hereafter) considers an asymptotic framework that takes $r \to \infty$. In the same spirit as IM, the MNW test is a Hausman-type test based on the variance of regression coefficients. Consider the full sample regression coefficient $\hat{\beta}$. Under the null hypothesis, the (full-sample) cluster-robust covariance estimator at the sub-cluster level, denoted, $\hat{\Omega}^\text{CCE}_J$, is consistent for the asymptotic variance-covariance matrix. 

Under either the null or the alternative, but maintaining the assumption that coarse clusters are independent, the (full-sample) cluster-robust covariance estimator at the cluster level, denoted, $\hat{\Omega}^\text{CCE}_K$, is consistent for the asymptotic variance-covariance matrix. Under the null hypothesis, the authors show that their test statistic converges to a standard normal distribution:
$
	\frac{\hat{\Omega}^\text{CCE}_K - \hat{\Omega}^\text{CCE}_J}{\hat{V}^{MNW}} \overset{d}{\to} N(0,1)
$
for an appropriately defined $\hat{V}^{MNW}$. 

It is well known that the cluster-robust covariance estimator can be severely biased when $r$ is small. In order to deal with such situations, the authors propose to conduct the test using wild (sub-)cluster bootstrap. They prove the consistency of this approach in their large-$r$ framework, showing power even against alternatives with negative correlations.  

Compared to the MNW test, our test is theoretically justified when both $r$ and $q$ are small, provided that $n_j$'s are large. Our test could therefore be preferable in such applications since it is presently not known if the MNW test remains valid once we take $r$ and $q$ to be fixed. However, as with the IM test, the MNW test allows unrestricted covariance within sub-clusters, whereas our test is expected to have poor performance if observations within sub-clusters are highly correlated. Our test is also conservative relative to the MNW test. On the other hand, simulation evidence suggests that it has comparable performance with the MNW test when clusters have differing variances (see Section \ref{section--montecarlo}). 



\section{Monte Carlo Simulations}\label{section--montecarlo}

In this section, we examine the finite sample performance of our worst-case randomization test (WCR) together with the IM and bootstrap version of the MNW tests via Monte Carlo simulations. We also study the performance of the na\"ive randomization test (NR) as described in Section \ref{subsubsection--naivetest}. 
We consider two data generating processes described below. 

\noindent \textbf{Model 1}: Model 1 is defined by the following:
\begin{gather*}
	Y_{t,j,k}  = X_{t,j,k}'\beta +  \sigma_{j,k}\left(\rho V_{t,k} + \frac{1}{\sqrt{1 - \phi^2}} U_{t,j,k}\right)~, \\
	V_{t,k}  \overset{\text{iid}}{\sim} N(0, 1) ~,	\quad U_{t,j,k}  = \phi U_{t-1,j,k} + \varepsilon_{t,j,k}, \quad \varepsilon_{t,j,k} \overset{\text{iid}}{\sim} N(0, 1) ~,
\end{gather*}
In particular, we set $X_{t,j,k} = \beta = 1$ and $\phi = 0.25$. Errors are correlated within a sub-cluster, according to an $AR(1)$ process, with autocorrelation coefficient $\phi$. $\rho$ captures the importance of cluster level shock. Since $\frac{1}{\sqrt{1 - \phi^2}} U_{t,j,k}$ has unit variance, $\rho$ is exactly the relative variance of cluster- to sub-cluster-level shocks. $\sigma_{j,k}$ controls the variance of the unobserved term in each cluster $k$. Here in Section \ref{subsection--mainsimulations}, we set $\sigma_{j,k} = 1$ for all $j \in \mathcal{J}, k \in \mathcal{K}$. In Section \ref{subsection--clusthet},  we explore the consequences of cluster heterogeneity by varying $\sigma_{j,k}$. 

\noindent \textbf{Model 2}: This is the model used in the simulations of \cite{mnw2020}, with the constant omitted. Let $m_k = \sum_{j \in \mathcal{J}_k} n_j$ be the total number of observations in cluster $k$. Let $U_k$ be the $m_k \times 1$ vector of $U_{t,j,k}$ for all observations in cluster $k$. Then
\begin{align*}
	U_k = \rho W_\xi\xi_k + \sqrt{1 - \rho^2} \, \epsilon_k \quad , \quad \epsilon_k \sim N(0, I_{m_k})~,
\end{align*}
where $\xi_k$ is a $10 \times 1$ vector distributed as:
\begin{align*}
	\xi_{k, 1} \sim N(0,1) \quad , \quad \xi_{k, l} = \phi\xi_{k, l-1} + e_{k,l} \quad , \quad e_{k,l} \sim N(0, 1- \phi^2) \quad , \quad l \in \{2,...,10\}
\end{align*}
and $W_\xi$ is the $m_k \times 10$ loading matrix with the $(i,j)^\text{th}$ entry $\mathbf{1}\left\{ j = \lfloor (i-1)10/m_k \rfloor + 1 \right\}$. Under this model, $\frac{1}{10}$ of the observations in each cluster are correlated because they depend directly on the same $\xi_{k, l}$. In addition, there is correlation between the $\xi_{k,l}$'s since it is generated according to an AR(1) process. Observations are then ordered so that every sub-cluster contains the same number of observations that depend on each $\xi_{k,l}$. Finally, $\beta = (1,1)'$ and the two covariates are independent and generated in the same way as $U$. This model features more complex correlations between and within the sub-clusters. Clusters are independent and identically distributed. As in Section 5.2 of \cite{mnw2020}, we set $\phi = 0.5$. $\rho$ here is directly comparable to $w_\xi$ in their simulations. 

For our simulations, we perform the test at the 5\% level. 1,000 Monte Carlo simulations were drawn for each combination of the parameters. The non-standard reference distribution in IM is evaluated using 1,000 Monte Carlo draws. Wild bootstrap in MNW is evaluated using $399$ draws as in their simulations. 


\subsection{Performance over values of $r$, $q_k$ and $n_j$}\label{subsection--mainsimulations}

To understand the size and power of each of our tests in scenarios with few clusters and few sub-clusters, we consider equal-sized clusters and sub-clusters, with $r \in \{4, 8, 12\}$, $q_k \in \{4, 8, 12\}$ and $n_j \in \{25, 50, 100\}$. We consider $\rho \in \{0, 0.5\}$.

Table \ref{table--mc_0_12} presents results under the null hypothesis ($\rho = 0$). Across the two models, we see that regardless of $r$, the IM test performs poorly when $q_k$ is small. With $q_k = 4$, type I error is between 15\%  and 20\%. By $q_k = 12$, however, the size is between 6-7\%.  Comparatively, our test, which is highly conservative, has type I error less than 2\% across all values of $q_k$. The MNW and NR tests perform well across the board. Table \ref{table--mc500_12}  presents results under the alternative $\rho = 0.5$. Relative to the IM and MNW tests, our test has power that is consistently lower. In particular, our test does poorly when $q_k$ is small. This is the weakness of the worst-case approach. 

\begin{table}[htbp]
  \centering \footnotesize
    \begin{tabular}{>{\centering}p{0.75cm} >{\centering}p{0.75cm} >{\centering}p{0.75cm} >{\centering}p{0.2cm} cccc >{\centering}p{0.2cm} cccc}
        \cmidrule{1-13}\morecmidrules\cmidrule{1-13}
          &       &       &       & \multicolumn{4}{c}{Model 1}   &       & \multicolumn{4}{c}{Model 2} \\
    \midrule
    $r$   & $q_k$ & $n_j$ &       & NR    & WCR   & IM    & MNW   &       & NR    & WCR   & IM    & MNW \\ \midrule
    \multirow{9}[5]{*}{4} & \multirow{3}[1]{*}{4} & 25    &       & 0.023 & 0.000 & 0.152 & 0.053 &       & 0.017 & 0.000 & 0.152 & 0.059 \\
          &       & 50    &       & 0.022 & 0.000 & 0.151 & 0.057 &       & 0.019 & 0.000 & 0.158 & 0.059 \\
          &       & 100   &       & 0.017 & 0.000 & 0.154 & 0.052 &       & 0.027 & 0.000 & 0.152 & 0.051 \\
\cmidrule{2-13}          & \multirow{3}[2]{*}{8} & 25    &       & 0.027 & 0.000 & 0.092 & 0.050 &       & 0.019 & 0.000 & 0.106 & 0.050 \\
          &       & 50    &       & 0.025 & 0.000 & 0.092 & 0.054 &       & 0.023 & 0.000 & 0.102 & 0.054 \\
          &       & 100   &       & 0.019 & 0.000 & 0.093 & 0.047 &       & 0.028 & 0.001 & 0.105 & 0.052 \\
\cmidrule{2-13}          & \multirow{3}[2]{*}{12} & 25    &       & 0.018 & 0.001 & 0.064 & 0.038 &       & 0.021 & 0.000 & 0.068 & 0.044 \\
          &       & 50    &       & 0.020 & 0.000 & 0.067 & 0.044 &       & 0.020 & 0.000 & 0.073 & 0.048 \\
          &       & 100   &       & 0.041 & 0.001 & 0.083 & 0.061 &       & 0.034 & 0.002 & 0.087 & 0.064 \\
    \midrule
    \multirow{9}[6]{*}{8} & \multirow{3}[2]{*}{4} & 25    &       & 0.020 & 0.000 & 0.171 & 0.063 &       & 0.022 & 0.000 & 0.169 & 0.048 \\
          &       & 50    &       & 0.017 & 0.000 & 0.153 & 0.054 &       & 0.026 & 0.000 & 0.170 & 0.049 \\
          &       & 100   &       & 0.035 & 0.000 & 0.189 & 0.067 &       & 0.023 & 0.000 & 0.159 & 0.050 \\
\cmidrule{2-13}          & \multirow{3}[2]{*}{8} & 25    &       & 0.022 & 0.000 & 0.088 & 0.052 &       & 0.026 & 0.000 & 0.109 & 0.065 \\
          &       & 50    &       & 0.029 & 0.001 & 0.089 & 0.048 &       & 0.028 & 0.001 & 0.097 & 0.051 \\
          &       & 100   &       & 0.025 & 0.001 & 0.101 & 0.058 &       & 0.025 & 0.002 & 0.101 & 0.052 \\
\cmidrule{2-13}          & \multirow{3}[2]{*}{12} & 25    &       & 0.027 & 0.000 & 0.078 & 0.050 &       & 0.027 & 0.000 & 0.072 & 0.049 \\
          &       & 50    &       & 0.024 & 0.001 & 0.060 & 0.037 &       & 0.033 & 0.001 & 0.093 & 0.053 \\
          &       & 100   &       & 0.021 & 0.000 & 0.068 & 0.043 &       & 0.026 & 0.001 & 0.076 & 0.037 \\
    \midrule
    \multirow{9}[6]{*}{12} & \multirow{3}[2]{*}{4} & 25    &       & 0.019 & 0.000 & 0.164 & 0.039 &       & 0.016 & 0.002 & 0.186 & 0.044 \\
          &       & 50    &       & 0.020 & 0.000 & 0.169 & 0.057 &       & 0.018 & 0.000 & 0.175 & 0.036 \\
          &       & 100   &       & 0.023 & 0.001 & 0.170 & 0.054 &       & 0.031 & 0.000 & 0.202 & 0.058 \\
\cmidrule{2-13}          & \multirow{3}[2]{*}{8} & 25    &       & 0.035 & 0.001 & 0.092 & 0.037 &       & 0.035 & 0.000 & 0.106 & 0.052 \\
          &       & 50    &       & 0.025 & 0.001 & 0.093 & 0.033 &       & 0.039 & 0.002 & 0.105 & 0.056 \\
          &       & 100   &       & 0.033 & 0.001 & 0.094 & 0.052 &       & 0.035 & 0.002 & 0.085 & 0.038 \\
\cmidrule{2-13}          & \multirow{3}[2]{*}{12} & 25    &       & 0.024 & 0.001 & 0.075 & 0.048 &       & 0.026 & 0.001 & 0.071 & 0.037 \\
          &       & 50    &       & 0.030 & 0.004 & 0.076 & 0.045 &       & 0.034 & 0.001 & 0.097 & 0.052 \\
          &       & 100   &       & 0.030 & 0.000 & 0.093 & 0.056 &       & 0.028 & 0.001 & 0.059 & 0.039 \\
        \cmidrule{1-13}\morecmidrules\cmidrule{1-13}
    \end{tabular}%
    \caption{Monte Carlo rejection rates under the null hypothesis $\rho = 0$ at 5\% level of significance. WCR refers to our worst-case randomization test. IM is the test from \cite{im2016}. MNW is the bootstrap version of the test in \cite{mnw2020}. NR is the na\"ive randomiztion test. $r$ is the number of clusters, $q_k$ is the number of sub-clusters in each cluster, and $n_j$ is the number of individuals in each sub-cluster.
    }  \label{table--mc_0_12}%
\end{table}%

\begin{table}[htbp]
  \centering \footnotesize 
    \begin{tabular}{>{\centering}p{0.75cm} >{\centering}p{0.75cm} >{\centering}p{0.75cm} >{\centering}p{0.2cm} cccc >{\centering}p{0.2cm} cccc}
    \cmidrule{1-13}\morecmidrules\cmidrule{1-13}
          &       &       &       & \multicolumn{4}{c}{Model 1}   &       & \multicolumn{4}{c}{Model 2} \\
    \midrule
    $r$   & $q_k$ & $n_j$ &       & NR    & WCR   & IM    & MNW   &       & NR    & WCR   & IM    & MNW \\ \midrule
    \multirow{9}[5]{*}{4} & \multirow{3}[1]{*}{4} & 25    &       & 0.053 & 0.000 & 0.305 & 0.155 &       & 0.089 & 0.000 & 0.380 & 0.205 \\
          &       & 50    &       & 0.071 & 0.000 & 0.345 & 0.164 &       & 0.112 & 0.000 & 0.493 & 0.312 \\
          &       & 100   &       & 0.052 & 0.000 & 0.301 & 0.153 &       & 0.240 & 0.003 & 0.672 & 0.544 \\
\cmidrule{2-13}          & \multirow{3}[2]{*}{8} & 25    &       & 0.098 & 0.000 & 0.364 & 0.276 &       & 0.167 & 0.010 & 0.472 & 0.380 \\
          &       & 50    &       & 0.108 & 0.003 & 0.354 & 0.262 &       & 0.346 & 0.043 & 0.639 & 0.585 \\
          &       & 100   &       & 0.093 & 0.001 & 0.364 & 0.270 &       & 0.511 & 0.130 & 0.791 & 0.751 \\
\cmidrule{2-13}          & \multirow{3}[2]{*}{12} & 25    &       & 0.176 & 0.029 & 0.467 & 0.403 &       & 0.249 & 0.059 & 0.524 & 0.480 \\
          &       & 50    &       & 0.190 & 0.024 & 0.471 & 0.405 &       & 0.460 & 0.190 & 0.742 & 0.697 \\
          &       & 100   &       & 0.197 & 0.032 & 0.495 & 0.424 &       & 0.657 & 0.332 & 0.861 & 0.848 \\
    \midrule
    \multirow{9}[6]{*}{8} & \multirow{3}[2]{*}{4} & 25    &       & 0.079 & 0.000 & 0.427 & 0.195 &       & 0.117 & 0.002 & 0.494 & 0.290 \\
          &       & 50    &       & 0.088 & 0.002 & 0.460 & 0.231 &       & 0.247 & 0.011 & 0.711 & 0.537 \\
          &       & 100   &       & 0.084 & 0.000 & 0.437 & 0.187 &       & 0.445 & 0.053 & 0.878 & 0.765 \\
\cmidrule{2-13}          & \multirow{3}[2]{*}{8} & 25    &       & 0.194 & 0.025 & 0.543 & 0.424 &       & 0.281 & 0.056 & 0.672 & 0.605 \\
          &       & 50    &       & 0.176 & 0.025 & 0.547 & 0.432 &       & 0.503 & 0.225 & 0.841 & 0.806 \\
          &       & 100   &       & 0.205 & 0.035 & 0.571 & 0.475 &       & 0.770 & 0.483 & 0.959 & 0.945 \\
\cmidrule{2-13}          & \multirow{3}[2]{*}{12} & 25    &       & 0.318 & 0.107 & 0.690 & 0.625 &       & 0.443 & 0.209 & 0.796 & 0.762 \\
          &       & 50    &       & 0.305 & 0.100 & 0.683 & 0.625 &       & 0.711 & 0.479 & 0.930 & 0.926 \\
          &       & 100   &       & 0.324 & 0.107 & 0.685 & 0.617 &       & 0.912 & 0.766 & 0.990 & 0.988 \\
    \midrule
    \multirow{9}[6]{*}{12} & \multirow{3}[2]{*}{4} & 25    &       & 0.099 & 0.001 & 0.517 & 0.242 &       & 0.143 & 0.007 & 0.609 & 0.373 \\
          &       & 50    &       & 0.100 & 0.003 & 0.534 & 0.261 &       & 0.271 & 0.034 & 0.808 & 0.626 \\
          &       & 100   &       & 0.097 & 0.003 & 0.516 & 0.246 &       & 0.519 & 0.128 & 0.964 & 0.886 \\
\cmidrule{2-13}          & \multirow{3}[2]{*}{8} & 25    &       & 0.274 & 0.073 & 0.663 & 0.540 &       & 0.433 & 0.130 & 0.788 & 0.745 \\
          &       & 50    &       & 0.293 & 0.067 & 0.688 & 0.573 &       & 0.695 & 0.402 & 0.939 & 0.926 \\
          &       & 100   &       & 0.275 & 0.056 & 0.672 & 0.542 &       & 0.907 & 0.762 & 0.997 & 0.993 \\
\cmidrule{2-13}          & \multirow{3}[2]{*}{12} & 25    &       & 0.432 & 0.204 & 0.828 & 0.768 &       & 0.631 & 0.382 & 0.915 & 0.893 \\
          &       & 50    &       & 0.440 & 0.191 & 0.809 & 0.746 &       & 0.866 & 0.689 & 0.987 & 0.988 \\
          &       & 100   &       & 0.426 & 0.185 & 0.830 & 0.786 &       & 0.971 & 0.922 & 0.998 & 0.996 \\
    \cmidrule{1-13}\morecmidrules\cmidrule{1-13}
    \end{tabular}%
      \caption{Monte Carlo rejection rates under the alternative hypothesis $\rho = 0.5$ at 5\% level of significance. WCR refers to our worst-case randomization test. IM is the test from \cite{im2016}. MNW is the bootstrap version of the test in \cite{mnw2020}. NR is the na\"ive randomiztion test. $r$ is the number of clusters, $q_k$ is the number of sub-clusters in each cluster, and $n_j$ is the number of individuals in each sub-cluster.
      }  \label{table--mc500_12}%
\end{table}%

\begin{figure}[htpb]
\centering
\includegraphics[width=1\linewidth]{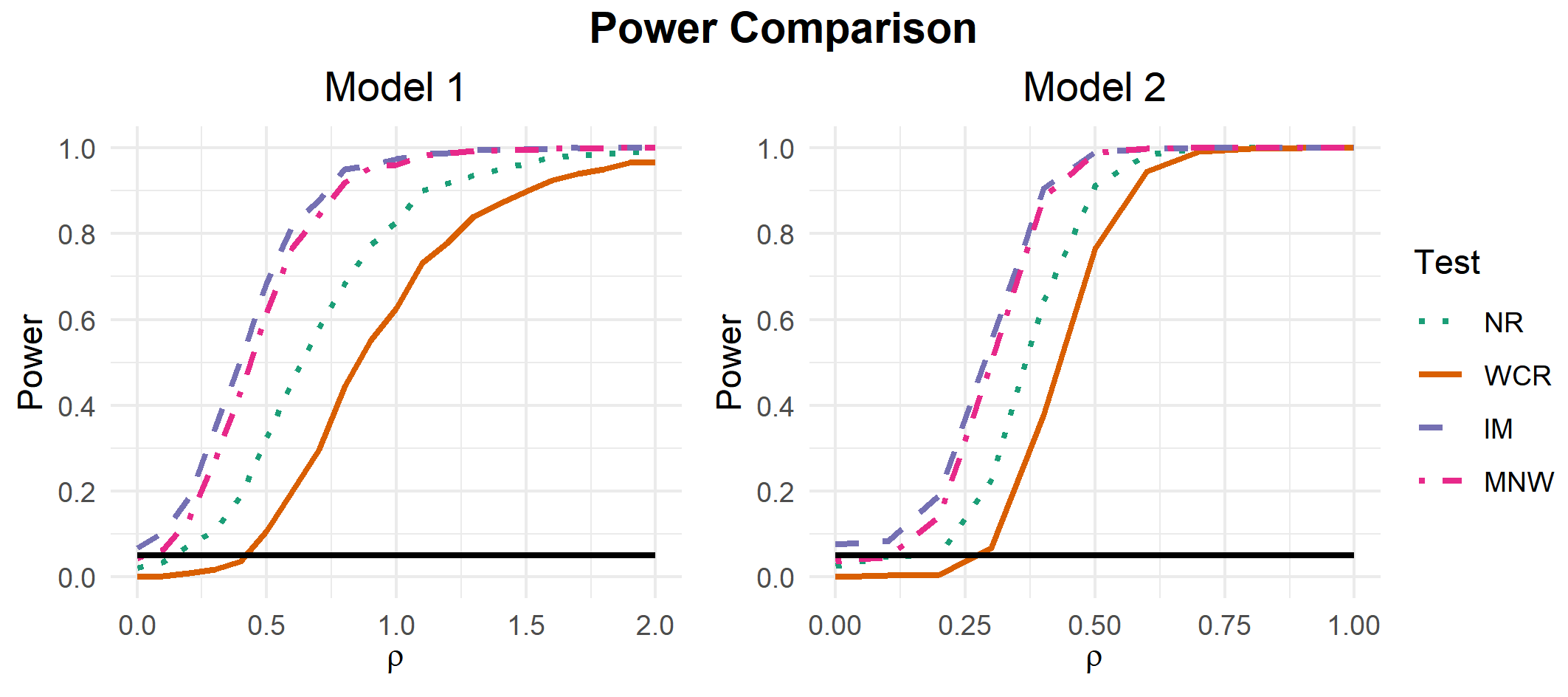}
\caption{Power of various tests for level of clustering when $r = 8$, $q_k = 8$, $n_j = 100$. The black line indicates the nominal size of the tests (5\%). 
}
\label{fig:powerseries12}
\end{figure}

Figure \ref{fig:powerseries12} presents power of the tests for $r = 8$, $q_k = 8$, $n_j = 100$ as we vary $\rho$ from $0$ to $2$ in model 1 and $0$ to $1$ in model 2. Across the two models, we see that the IM and MNW tests have greater power than our test. However, as $\rho$ increases, our test quickly catches up in power. 



\subsection{Effect of Cluster-Level Heterogeneity}\label{subsection--clusthet}

The previous section suggests that our test has poor performance compared to all other tests, including NR. However, a different picture emerges once we allow clusters and sub-clusters to be heterogeneous in their variances. 

\begin{figure}[htbp]
\centering
\includegraphics[width=.85\linewidth]{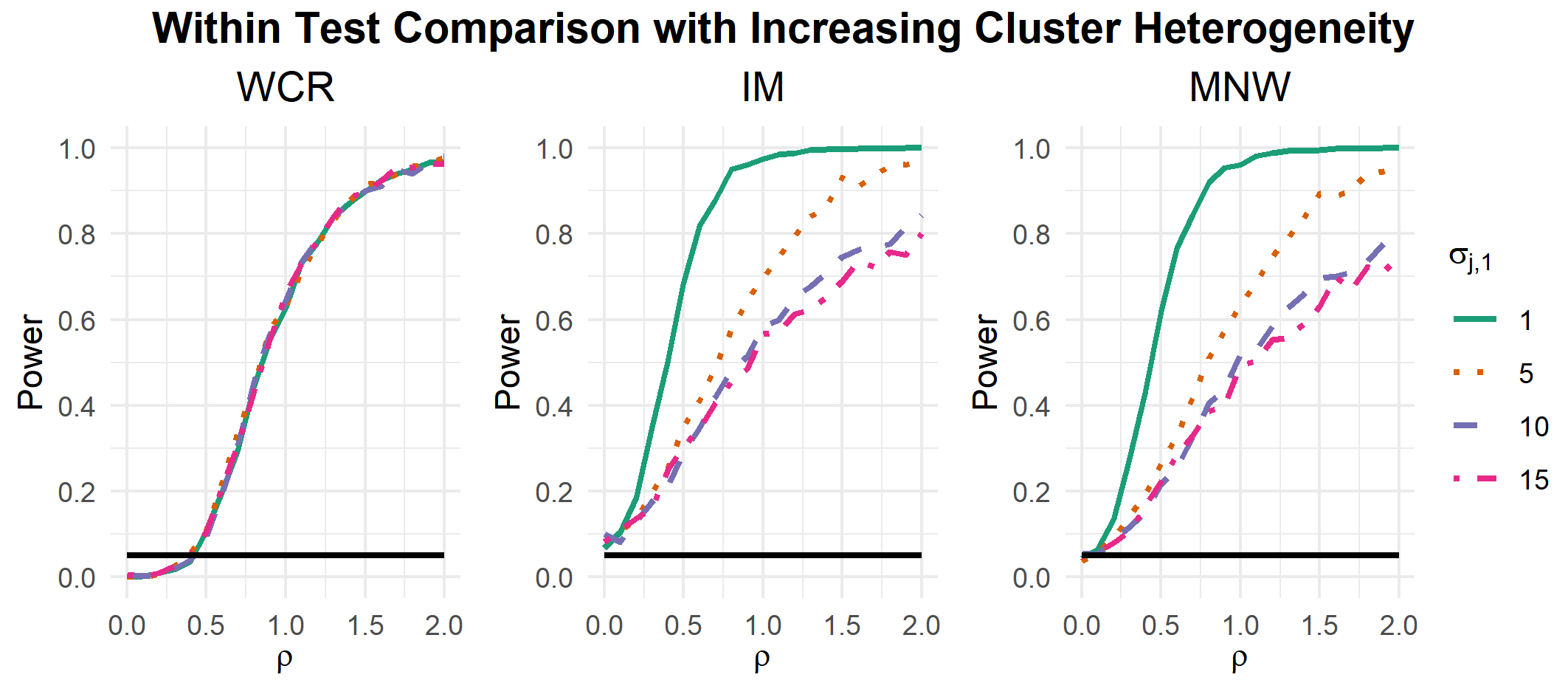}
\includegraphics[width=.85\linewidth]{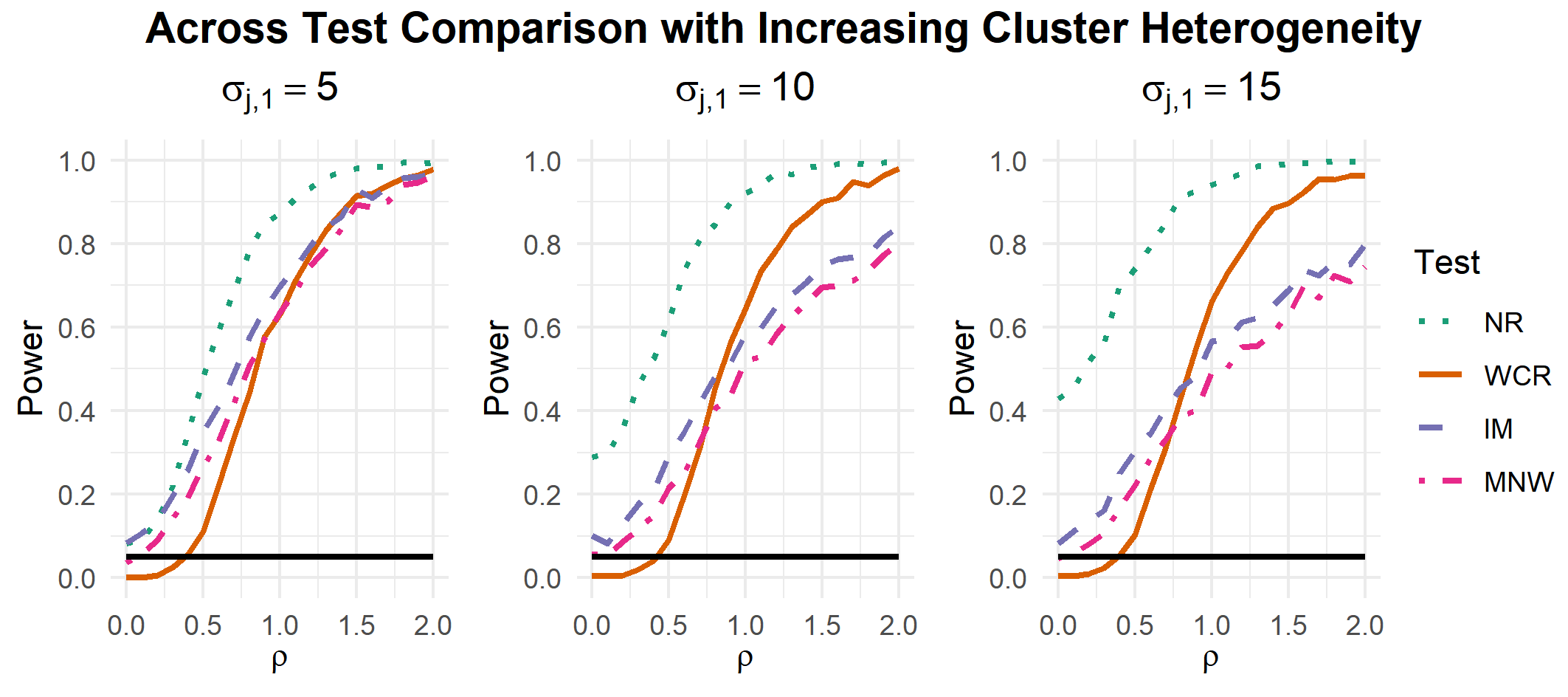}
\caption{Power of various tests for level of clustering in Model 1 as $\sigma_{j,1}$ increases. Here, $r = 8$, $q_k = 8$, $n_j = 100$. 
The black line indicates the nominal size of the tests (5\%).}
\label{fig:clusthet_across}
\end{figure}

We first consider what happens when clusters are heterogeneous. Specifically, we return to model 1 but with $\sigma_{j,1} \in \{5,10,15\}$. That is, when all sub-clusters in cluster 1 are much noisier than the rest. Figure \ref{fig:clusthet_across} plots power curves with $r = 8, q_k = 12, n_j = 100$ for $\sigma_{j,1} \in \{5,10,15\}$. These curves are directly comparable with Figure \ref{fig:powerseries12}. Starting from the within test comparison, we see that the performance of our test is unaffected by $\sigma_{j,1}$. However, power of IM and MNW quickly degrade as $\sigma_{j,1}$ increases. Turning to the across test comparison, we see that the tests perform similarly when $\sigma_{j,1} = 5$. As $\sigma_{j,1}$ increases to 10, our test starts to have more power than the IM and MNW tests for $\rho \geq 1$. The across test comparison also shows how the NR test fails to control size. In particular, when $\sigma_{j,1}$, an NR test with nominal size 5\% could wrongly reject over 40\% of the time. 

\begin{figure}[htbp]
\centering
\includegraphics[width=.85\linewidth]{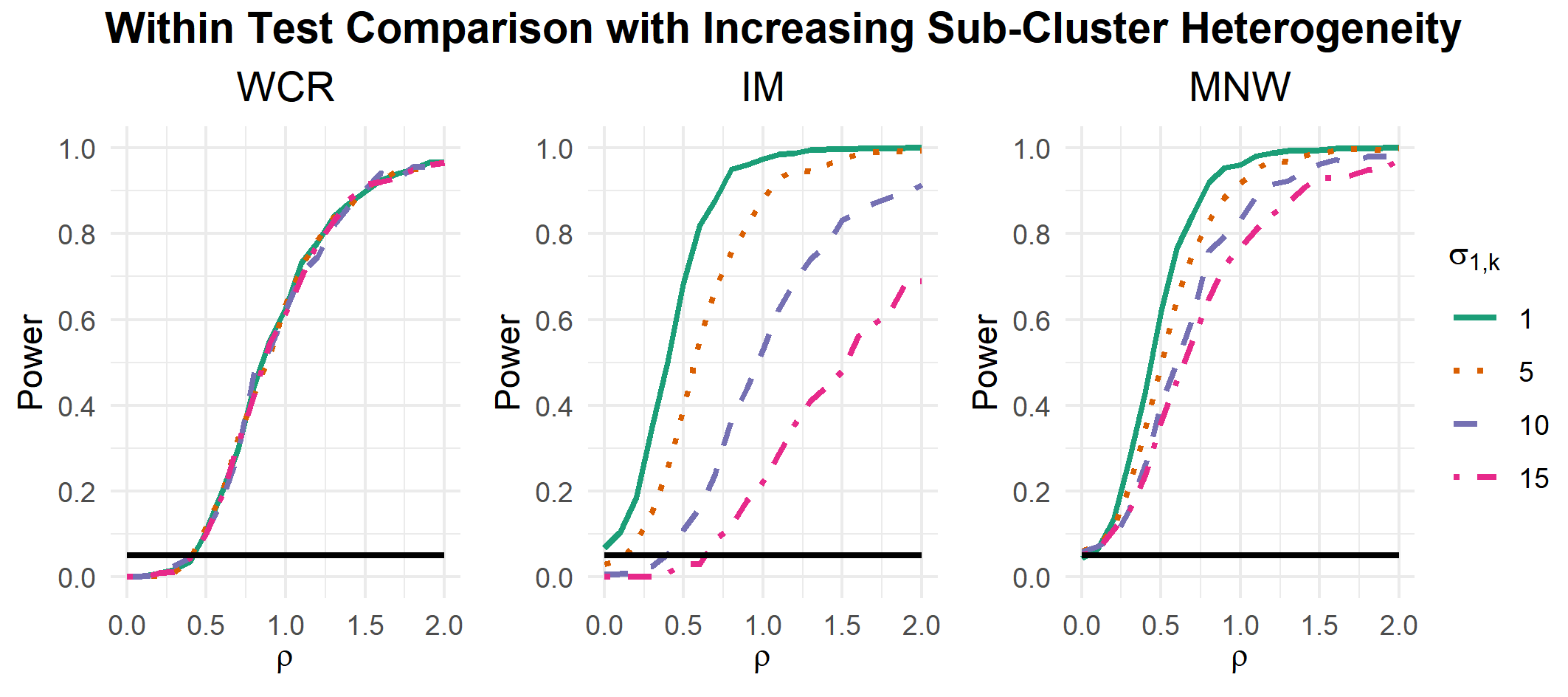}
\caption{Power of various tests for level of clustering in Model 1 as $\sigma_{1,k}$ increases. Here, $r = 8$, $q_k = 8$, $n_j = 100$. 
The black line indicates the nominal size of the tests (5\%).}
\label{fig:subclusthet_within}
\end{figure}

We see the same patterns when sub-clusters are heterogeneous. Consider again model 1 but with $\sigma_{1,k} \in \{5,10,15\}$. That is, when the first sub-cluster in each cluster is much noisier than the rest. Figure \ref{fig:subclusthet_within} presents the results. Again, our test is not affected by changing $\sigma_{1,k}$. The power of the IM test falls by a large extent as $\sigma_{1,k}$ increases. The MNW test is also negatively affected by $\sigma_{1,k}$, though less so than the IM test. 

All in all, the simulation evidence suggests that our test manages to maintain type I error below $\alpha$ when $q$ is small, whereas the IM and NR tests may see size distortion in such a setting. The cost of size control in a fixed $q$ setting is that the procedure is very conservative. This conservativeness limits the power of our test. However, the performance of our test is less sensitive to heterogeneous variances within and across clusters, such that it could be more powerful than the IM and MNW tests when some clusters or sub-clusters are much noisier than others. Hence, our test is suited for applications with small $q$ and heterogenous clusters. Indeed, as we will see in the next section, our test detects dependence in the clusters of \cite{gllqsx2019}, demonstrating its potential relevance for empirical work.


\section{Application: \cite{gllqsx2019}}\label{section--application}


In recent years, the poor performance of American students in assessment tests such as the Programme for International Student Assessment (PISA) has raised concerns among policymakers. \cite{gllqsx2019} argues that the testing gap reflects, among other things, the low effort that American students put in on tests, especially when compared to their higher scoring counterparts in other countries. 

The authors test their hypothesis by a randomized controlled experiment in which students were rewarded with cash for correct answers in a 25-question test. Those assigned to the treatment group were offered roughly \$1 USD per correct answer, while the control group received no payment. Students were informed right before the test started to prevent them from changing their effort in test preparation. The experiments were conducted at 4 schools in Shanghai and 2 schools in the US. Due to logistical reasons, the authors randomized treatment at the class level for some schools and individual level for others. 

Various regression analyses were conducted to study the effect of treatment on test-taking effort and test performance. Panel A in Table 3 examines whether monetary incentive increased the probability that students attempt a given question -- a proxy for effort.  It does so by estimating the following equation:
\begin{equation*}
	Y_{qi} = \beta Z_i + \gamma'W_i + \epsilon_{qi}~.
\end{equation*}
Here, the unit of analysis is a question and $Y_{qi}$ is an indicator for whether student $i$ attempted question $q$. $Z_i$ is the treatment indicator and $W_i$ is a vector of control variables, which include terms such as gender, ethnicity as well as question number fixed effects. We focus on Column 1 in Panel A, which looks at US students' responses to all 25 questions in the test, and Column 4, which looks at Shanghai students' responses to the same test. 

The authors present their linear regression estimate of $\beta$, together with standard errors clustered at the level of randomization. However, other levels of clustering are plausible:
\begin{itemize}
	\item G: Group Level, that is, the level of randomization.
	\item S: School Level.
	\item SY: Experiments in Shanghai schools were conducted in 2016 and then 2018. We could plausibly interact school and year of experiment. 
	\item ST: Schools in the US separate students into tracks (Honors, Regular, Others). We could plausibly interact school and track. 
\end{itemize}
We will refer to these levels of clustering by their initials hereafter. More information on the sizes of clusters can be found in appendix \ref{appendix--clusterstatistics}.

While the authors chose to cluster their standard errors by $G$, it seems reasonable to be concerned about correlation across individuals within the same school or among those who took the test in the same year. If these clusters were not independent, $t$-tests using the presented standard errors could lead to the wrong conclusions.

\begin{table}[h!]
  \centering \small
    \begin{tabular}{lcccc >{\centering}p{0.5cm} ccc}
     \cmidrule{1-9}\morecmidrules\cmidrule{1-9}        
          &       & \multicolumn{3}{c}{Column 1} &       & \multicolumn{3}{c}{Column 4} \\
    \cmidrule{3-5}\cmidrule{7-9}   
    $\hat{\beta}$ &       & \multicolumn{3}{c}{0.037} &       & \multicolumn{3}{c}{-0.030} \\[6pt]
\cmidrule{1-9}    

          &       & G     & ST    & S     &       & G     & SY    & S \\
    \midrule
    CCE S.E. &       & 0.017 & 0.008 & 0.000 &       & 0.008 & 0.020 & 0.023 \\
    CCE $p$ &       & 0.029 & 0.000 & 0.000 &       & 0.000 & 0.131 & 0.188 \\
    Wild Bootstrap $p$ &       & 0.064 & 0.073 & 0.262 &       & 0.002 & 0.152 & 0.126 \\
    ART $p$ &       & -     & 0.063 & 0.500 &       & -     & 0.125 & 0.250 \\
    IM2010 $p$ &       & -     & 0.926 & 0.974 &       & -     & 0.748 & 0.816 \\
     \cmidrule{1-9}\morecmidrules\cmidrule{1-9}        
    \end{tabular}%
  \caption{Tests for $\beta = 0$ under various levels of clustering. Based on the regressions in Table 3 Panel A of \cite{gllqsx2019}.}
  \label{tab:application-potentialresults}%
\end{table}%

Table \ref{tab:application-potentialresults} presents the OLS estimates from \cite{gllqsx2019} as well as the $p$-values that would be obtained from testing the null hypothesis that $\beta = 0$ using several methods. Specifically, we consider the wild cluster bootstrap (\cite{cgm2008}), approximate randomization tests (\cite{crs2017}) and the t-distribution based procedure of \cite{ibragimov2010t}, denoted IM2010. We perform these tests using the various plausible levels of clustering. For Column 1, we consider the increasingly coarse levels of clustering $G$, $ST$ and $S$. For the US, there are no schools sampled over multiple years, so $SY$ is the same as $S$. For Column 4, we consider the increasingly coarse levels of clustering $G$, $SY$ and $S$. In Shanghai schools, students are not separated by track, so $ST$ is the same as $S$. 

\begin{remark}
	\cite{gllqsx2019} present clustered standard errors but do not use them for inference. Instead, they conduct randomization inference by permuting treatment status as in \cite{young2019channeling}. This procedure tests the null hypothesis that the distribution of the $Y_{qi}$'s are the same with and without treatment. This is a stronger null hypothesis than the null of $0$ average treatment effect ($\beta = 0$). We believe that the latter hypothesis is typically the one of interest and test it in our Table \ref{tab:application-potentialresults}. 
\end{remark}

Turning to the results, for column 1, we see that CCE SE's decrease as we move to increasingly coarse levels of clustering. Correspondingly, $p$-values from CCE-based $t$-tests decrease as we coarsen the clusters. Such a pattern is typically interpreted as arising from the downward bias of CCEs with few clusters (\cite{mhe2008}), so that these $p$-values would be considered unreliable. Faced with downward bias, practitioners commonly turn to the wild cluster bootstrap. With this method, the $p$-values increase as we coarsen the clusters. While clustering at $G$ and $ST$ may lead one to conclude that there is strong evidence that $\beta \neq 0$, the $p$-value at $S$ suggests the absence of strong evidence. The same phenomenon arises with approximate randomization tests: at $ST$ there appears to be strong evidence that $\beta \neq 0$. At $S$, this is no longer true. With IM2010, the test does not reject in either case. We note that ART and IM2010 cannot be applied with $G$ as the chosen level of clustering, since both methods require $\beta$ to be estimated cluster-by-cluster. The results for column 4 are qualitatively similar. At $G$, CCE-based $t$-test and the wild cluster bootstrap find strong evidence that $\beta \neq 0$. This conclusion is overturned once we cluster at either $SY$ or $S$. 

\begin{table}[h!]
  \centering \small 
    \begin{tabular}{lcccc >{\centering}p{0.5cm} ccc}
     \cmidrule{1-9}\morecmidrules\cmidrule{1-9}    
          &       & \multicolumn{3}{c}{Column 1} &       & \multicolumn{3}{c}{Column 4} \\
\cmidrule{3-5}\cmidrule{7-9}          &       & G $\to$ ST & G $\to$ S & ST $\to$ S &       & G $\to$ SY & G $\to$ S & SY $\to$ S \\
    \midrule
    WCR   &       & 0.891 & 1.000 & 1.000 &       & 0.062 & 0.056 & 1.000 \\
    IM    &       & 0.817 & 0.868 & 0.673 &       & 0.000 & 0.006 & 0.019 \\
    MNW &       & 0.266 & 0.228 & 0.145 &       & 0.000 & 0.003 & 0.624 \\
     \cmidrule{1-9}\morecmidrules\cmidrule{1-9}        
    \end{tabular}%
  \caption{Tests of levels of clustering applied to the regression in Table 3 Panel A of \cite{gllqsx2019}.}    
  \label{tab:application-clustertests}%
\end{table}%

To assess the validity of the above specifications, we apply our WCR test, the IM test and the MNW tests. Table \ref{tab:application-clustertests} presents the resulting $p$-values. The notation $G \to S$ means that the null hypothesis involves sub-clusters $G$ and coarse clusters $S$. For Column 1, clustering at $G$ appears to be appropriate, as all 3 tests fail to reject the null hypotheses $G \to ST$ and $G \to S$. For Column 4, all 3 tests find strong evidence that sub-clusters $G$ are inappropriate. The WCR test has higher $p$-values than the IM and MNW tests, likely due to its lower power. Nonetheless, they are close to 5\%. The WCR and MNW tests do not reject the null hypothesis for $SY \to S$, whereas the IM test does. Given the that there are at most 2 school$\times$year per school, the IM test is likely to over-reject. As such, we consider the conclusion of the WCR and MNW test to be more reliable in this instance. Thus, results based on clustering at $SY$ are plausible.

All in all, we see that settings with varying numbers of clusters and sub-clusters arise in empirical work. Our test, designed for applications with few clusters and sub-clusters is relevant and appears to work well in practical settings. 

 
\section{Conclusion}\label{section--conclusion}
We propose to test for the level of clustering in a regression by means of a modified randomization test. We show that the test controls size even when the number of clusters and sub-clusters are small, provided that the size of sub-clusters are relatively large. This is a challenging situation not accommodated by existing tests. To ensure size control, our procedure may be conservative when clusters are homogeneous. However, in settings with heterogeneous clusters, it has power that is comparable with other tests. As such, our test can be useful when the researcher faces an application with few sub-clusters, particularly when these clusters are likely to be heterogeneous. Finally, we note that the test is easy to implement and could serve as a helpful robustness check to researchers working with clustered data. An R package is available from the author's website. 

\begin{center}
{\large\bf SUPPLEMENTARY MATERIAL}
\end{center}

\begin{description}

\item[Extended Appendices] Technical details including proof of theorem 1, details concerning the application and additional Monte Carlo simulations.

\end{description}


\bibliographystyle{chicago}
\bibliography{clusters}

\clearpage

\appendix

\begin{center}
{\large\bf SUPPLEMENTARY MATERIAL}
\end{center}

\section*{Appendices}

\section{Proof for Theorem \ref{proposition--size}}\label{appendix--proofs}

We first write:
\begin{align}
	& \frac{1}{\sqrt{n_j}} \sum_{i \in \mathcal{I}_j} \left(X_i - W_i\hat{\Pi}_j\right)\hat{U}_i   \nonumber \\
	= & \frac{1}{\sqrt{n_j}} \sum_{i \in \mathcal{I}_j} \left(X_i - W_i\hat{\Pi}_j\right){U}_i \label{prop1--partA}\\
	& - \frac{1}{\sqrt{n_j}} \sum_{i \in \mathcal{I}_j} \left(X_i - W_i\hat{\Pi}_j\right)^2\left(\hat{\beta} - \beta  \right) \label{prop1--partB}\\
	& - \frac{1}{\sqrt{n_j}} \sum_{i \in \mathcal{I}_j} \left(X_i - W_i'\hat{\Pi}_j\right)\left(W_i'\hat{\Pi}_j(\hat{\beta}-\beta) + W_i'(\hat{\gamma} - \gamma)  \right)~. \label{prop1--partC}
\end{align}
We analyse parts (\ref{prop1--partA}) and (\ref{prop1--partC}) in turn. Part (\ref{prop1--partB}) is handled using our worst case bound. Starting with (\ref{prop1--partA}):
\begin{align*}
	& \frac{1}{\sqrt{n_j}} \sum_{i \in \mathcal{I}_j} \left(X_i - W_i\hat{\Pi}_j\right){U}_i \\
	& = \frac{1}{\sqrt{n_j}} \sum_{i \in \mathcal{I}_j} \left(X_i - W_i{\Pi}_j\right){U}_i - \left(\hat{\Pi}_j - \Pi_j \right) \frac{1}{\sqrt{n_j}} \sum_{i \in \mathcal{I}_j} W_i{U}_i \\
	& = \text{N}\left({0}, \sigma_j^2 \right) + o_p(1)O_p(1)~,
\end{align*}
where the last equation follows from assumption \ref{assumption--clt}. Next for term (\ref{prop1--partC}), note that:
\begin{align*}
	& \frac{1}{\sqrt{n_j}} \sum_{i \in \mathcal{I}_j} W_i\left(X_i - W_i'\hat{\Pi}_j\right) \\
	& = \frac{1}{\sqrt{n_j}} \sum_{i \in \mathcal{I}_j} W_i\left(X_i - W_i{\Pi}_j\right) - \sqrt{n_j}\left(\hat{\Pi}_j - \Pi_j\right) \frac{1}{{n_j}} \sum_{i \in \mathcal{I}_j} W_iW_i' 	=  O_p(1)~.
\end{align*}
As such, by assumption (\ref{assumption--clt}),
\begin{align}\label{equation--tilderesidualisop1}
	\left(\hat{\Pi}_j'(\hat{\beta}-\beta) + (\hat{\gamma} - \gamma)  \right)' \frac{1}{\sqrt{n_j}} \sum_{i \in \mathcal{I}_j} W_i\left(X_i - W_i'\hat{\Pi}_j\right) = o_p(1)~.
\end{align}
Our analysis of terms (\ref{prop1--partA}) and (\ref{prop1--partC}) show that setting $\lambda = \hat{\beta} - \beta$, we have that
\begin{equation*}
\hat{S}_{n}(\hat{\beta} - \beta) \overset{d}{\to} \text{N}(\mathbf{0}, \Sigma)~.
\end{equation*}
Furthermore, under the null hypothesis, $\Sigma$ is diagonal since $$\frac{1}{\sqrt{n_j}} \sum_{i \in \mathcal{I}_j} \left(X_i - W_i{\Pi}_j\right){U}_i$$
is independent across sub-clusters. 

Maintaining the assumption that $\hat{\beta} - \beta$ is known, we show that the requirements for Theorem 3.1 in \cite{crs2017} are met. 
\begin{itemize}
	\item[i.] $\hat{S}_{n}(\hat{\beta} - \beta) \overset{d}{\to} S$ by the analysis above. 
	\item[ii.] By symmetry of $S$ about $0$ and the fact that $\Sigma$ is diagonal, it is immediate that $gS$ has the same distribution as $S$ under the null hypothesis.
	\item[iii.] For all $g \neq g'$, $P(T(gS) \neq T(g'S)) = 1$. This is because for a given component $j$ on which $g_j \neq g'_j$, $S_j = -S_j$ if and only if $S_j = 0$, which occurs with probability 0. 
\end{itemize}
Hence, we have that:
\begin{equation*}
	E\left[\mathbf{1} \left\{  p(\hat{S}_n(\hat{\beta} - \beta)) \leq \alpha  \right\}     \right] \to \alpha~.
\end{equation*}
This test is conservative since we break ties in favour of not rejecting the null-hypothesis. It is then immediate that:
\begin{equation*}
\pushQED{\qed} 
	\underset{n \to \infty}{\lim \sup} \, E\left[\mathbf{1} \left\{ \sup_{\lambda \in \mathbf{R}} p(\hat{S}_n(\lambda)) \leq \alpha  \right\}     \right] \leq \underset{n \to \infty}{\lim \sup} \, E\left[\mathbf{1} \left\{  p(\hat{S}_n(\hat{\beta} - \beta)) \leq \alpha  \right\} \right] = \alpha~. \qedhere
\popQED 
\end{equation*}

\section{Inference with Unnecessarily Coarse Clusters}\label{appendix--inferecefewclusters}

In this section, we demonstrate by simulation the problems with using unnecessarily coarse clusters for inference. Consider the model:
\begin{gather*}
	Y_{t,j,k}  = \beta + \frac{1}{\sqrt{1 - \phi^2}} U_{t,j,k}~, \\
	U_{t,j,k}  = \phi U_{t-1,j,k} + \varepsilon_{t,j,k}, \quad \varepsilon_{t,j,k} \overset{\text{iid}}{\sim} N(0, 1), \quad U_{1,j,k} \overset{\text{iid}}{\sim} N(0,1)
\end{gather*}
where $t$ is an observation from fine cluster $j$ in coarse cluster $k$. Here, individuals in the same fine cluster $j$ are dependent due to $U_{t,j,k}$, but individuals in different fine clusters are independent. Clustering at both the fine and coarse levels are therefore valid, though as we show, unncessarily coarse clustering will lead to issues. Suppose there are 4 coarse clusters, 12 fine clusters in each coarse clusters, and 100 observations per fine cluster. Further set $\phi = 0.25$ and $\beta = 1$. Suppose want to test the hypothesis $\beta = \beta_0$ at 5\% level of significance. The most popular options are CCE-based tests, ARTs, or wild bootstrap based tests, all of which can be implemented with clustering at either the $k$ level, or at the $j$ level. 

\begin{figure}[htpb]
\centering
\includegraphics[width=1\linewidth]{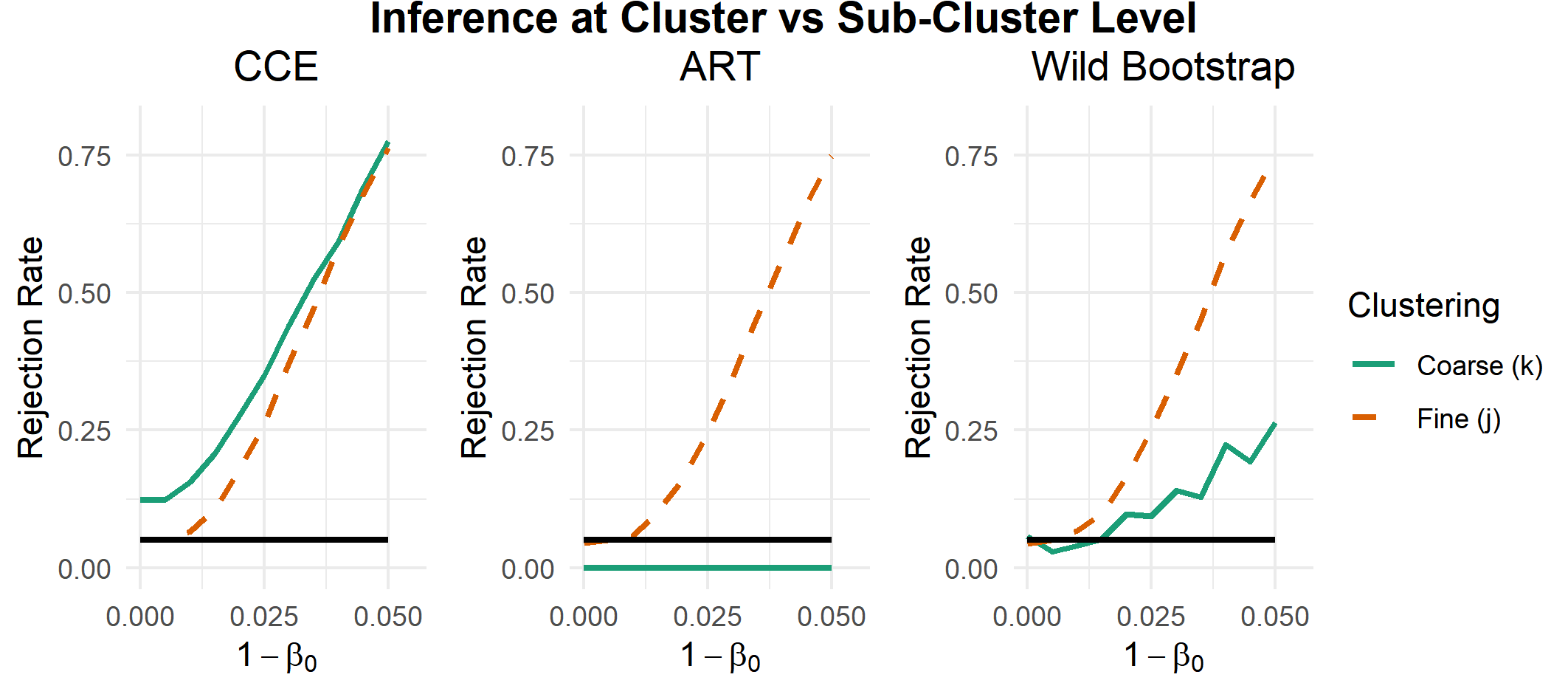}
\caption{Rejection rates from implementing CCE-based tests, ARTs or wild bootstrap-based tests in a simple model (see text). We assume 4 coarse clusters, 12 fine clusters per coarse clusters and 100 observations per fine cluster. Simulation are repeated 1000 times. The black line indicates the nominal size of the test (5\%).}
\label{fig:powersize}
\end{figure}

Figure \ref{fig:powersize} presents the rejection rates from each of these tests as we vary $\beta_0$ from 1 to 0 (equivalently, as $1-\beta_0$ varies from 0 to 1). The left panel pertains to CCE-based tests. Under the null hypothesis, when $\beta_0 = 1$ (i.e when $1-\beta_0 = 0$), the test using coarse clustering rejects over 12\% of the time, more than twice the nominal size. However, same test controls size well with fine clustering. The middle panel presents results from conservative ARTs that do not perform random tie-breaking. The test controls size and has good power with fine clustering. With coarse clustering, however, the test never rejects. This is because conservative ARTs can reject only when the size is smaller than $2^{q-1}$, where $q$ is the number of clusters. This is an extreme example of how using coarse clusters could dramatically reduce power. The right panel presents results from wild bootstrap-based tests (\cite{cgm2008}) commonly used when the number of clusters is small. Our setting satisfies the requirements of \cite{css2019}, so that we expect tests based on either level of clustering to control size, as they indeed do. However, the test with coarse clustering has much lower power than with fine clustering. All in all, our simulation shows that inference using unnecessarily coarse levels of clustering leads to problems. 

\section{Residualized Null Hypothesis}\label{section--residualizednull}

In this section, we explain why a researcher conducting inference on $\beta$ only has to test the residualized hypothesis in equation (\ref{equation--nullhypothesis}). As in standard notation, let $W$ be the matrix with $W_i'$ in its $i^\text{th}$ row. In the following we assume that $W$ has full rank for convenience.\footnote{The case where $W$ is rank deficient is similar except $\hat{\Pi}$ might not have an explicit formula. However, this does not change the properties of the projection residuals.} Write:
\begin{align*}\small
	P_W Y & = W(W'W)^{-1}W'Y \\
		& = W(W'W)^{-1}W'\left(X\hat{\beta} + W'\hat{\gamma} + \hat{U} \right) \\
		& = W\hat{\gamma} + W\hat{\Pi}\hat{\beta}~,
\end{align*}
where $\hat{\beta}$ and $\hat{\gamma}$ are full sample OLS estimates from equation (\ref{equation--model}) and $\hat{U}$ are the residuals from the same regression. $\hat{\Pi} = (W'W)^{-1}W'X$ is one of the possible consistent estimators for $\Pi$ when $W$ has full rank. By the Frisch-Waugh-Lovell theorem, we can then write:
\begin{align*}\small 
	\hat{\beta} & = \left( \left(X - P_W X\right)'\left(X - P_W X\right)  \right)^{-1} \left(X - P_W X\right)'\left(Y - P_W Y\right) \\
	& = \frac{\sum_{i=1}^n  \left(X_i - W_i \hat{\Pi} \right)\left(Y_i - W_i \hat{\gamma} - W_i \hat{\Pi}\hat{\beta} \right)    }{\sum_{i=1}^n  \left(X_i - W_i \hat{\Pi} \right)^2} \\
	& = \frac{\sum_{i=1}^n  \left(X_i - W_i \hat{\Pi} \right)\left( (X_i -W_i\hat\Pi)\beta + W_i\hat{\Pi}\beta + W_i\gamma +  U_i  - W_i \hat{\gamma} - W_i \hat{\Pi}\hat{\beta} \right)    }{\sum_{i=1}^n  \left(X_i - W_i \hat{\Pi} \right)^2} \\
	& = \beta +  \frac{\sum_{i=1}^n  \left(X_i - W_i \hat{\Pi} \right)\left(U_i - W_i\hat{\Pi}(\hat{\beta} - \beta) -  W_i (\hat{\gamma} - \gamma) \right)    }{\sum_{i=1}^n  \left(X_i - W_i \hat{\Pi} \right)^2} ~.
\end{align*}
Then, by the same argument that leads to equation (\ref{equation--tilderesidualisop1}), we have that under consistency of $\hat{\gamma}$, $\hat{\beta}$ and $\sqrt{n}$-consistency of $\hat{\Pi}$:
\begin{equation*}
	\sqrt{n}\left(\hat{\beta} - \beta\right) = \frac{\frac{1}{\sqrt{n}}\sum_{i=1}^n  \left(X_i - W_i{\Pi} \right)U_i  }{\frac{1}{n}\sum_{i=1}^n  \left(X_i - W_i{\Pi} \right)^2} +o_p(1)~.
\end{equation*}
Or, if we are estimating $\hat{\beta}$ cluster-by-cluster, 
\begin{equation*}
	\sqrt{n_j}\left(\hat{\beta}_j - \beta\right) = \frac{\frac{1}{\sqrt{n_j}}\sum_{i=1}^{n_j}  \left(X_i - W_i{\Pi}_j \right)U_i  }{\frac{1}{n_j}\sum_{i=1}^n  \left(X_i - W_i{\Pi}_j \right)^2} +o_p(1)~.
\end{equation*}
As such, the asymptotic distribution of the $\hat{\beta}$/$\hat{\beta}_j$'s depend only on $Z_i = \left(X_i - W_i{\Pi} \right)U_i$. If there is no dependence in $Z_i$ across the sub-clusters, approximate randomization test at the sub-cluster level yields valid inference. 

Our asymptotic framework takes the number of sub-clusters as fixed. However, if there is no dependence across sub-clusters in the $Z_i$'s, then provided that the usual regularity conditions hold (e.g. in \cite{hl2019}), we have that as the number of sub-cluster grows to infinity, inference with CCE clustered at the sub-cluster level also leads to correct inference. 

\section{Restricted Heterogeneity implied by Assumption 2}\label{section--impliedheterogeneity}

In this section we explain how the assumption that $S_n \to N(0, \Sigma)$ requires that $\min_j n_j \to \infty$, but does not place any restrictions on the relative rates at which each $n_j \to \infty$. We do this by way of an example. Assume that each $S_{n,j}$ have $2 + \delta$ moments for $\delta > 0$ and is weakly dependent in the sense of \cite{doukhan1999new}. This is a general form of dependence which includes strongly mixing sequences and Bernoulli shifts as special cases. \cite{nze2004weak} argues for usefulness in econometrics. We show that under our assumption, $S_n \to N(0, \Sigma)$ as long as $\min_j n_j \to \infty$, even if $\frac{n_{j'}}{\min_j n_j} \to \infty$ for some $j' \in [q]$. 

Consider the Cramer-Wold theorem, which gives us that $S_n \overset{d}{\to} N(0, \Sigma)$ under the null hypothesis if and only if for all $\lambda \in \mathbf{R}^q$ we have
\begin{align*}
	E\left[\exp\left(it\lambda'S_n\right)\right]
	= \prod_{j=1}^q E\left[\exp\left(it\lambda_jS_{n,j}\right)\right] \to \prod_{j=1}^q E\left[\exp\left(-it\frac{\lambda_j^2\sigma_j^2 t^2}{2}\right)\right]~.
\end{align*}
Since characteristic functions are bounded by 1, 
\begin{gather*}
	\left\lvert \prod_{j=1}^q E\left[\exp\left(it\lambda_jS_{n,j}\right)\right] - \prod_{j=1}^q E\left[\exp\left(-it\frac{\lambda_j^2\sigma_j^2 t^2}{2}\right)\right] \right\rvert
	\leq \sum_{j = 1}^q \left\lvert  E\left[\exp\left(it\lambda_jS_{n,j}\right)\right] -   E\left[\exp\left(-it\frac{\lambda_j^2\sigma_j^2 t^2}{2}\right)\right] \right\rvert~.
\end{gather*}
Proposition 7.1 in \cite{dedecker2007weak} yields:
\begin{equation*}
	\left\lvert  E\left[\exp\left(it\lambda_jS_{n,j}\right)\right] -   E\left[\exp\left(-it\frac{\lambda_j^2\sigma_j^2 t^2}{2}\right)\right] \right\rvert \leq Cn^{-c_j^*}
\end{equation*}
where $c_j^* > 0$ depends on the amount of dependence within sub-cluster $j$. Define $c = \min_j c_j$. Then we can write
\begin{equation*}
	\left\lvert \prod_{j=1}^q E\left[\exp\left(it\lambda_jS_{n,j}\right)\right] - \prod_{j=1}^q E\left[\exp\left(-it\frac{\lambda_j^2\sigma_j^2 t^2}{2}\right)\right] \right\rvert = O\left( \left(\min_{j \in [q]} n_j\right)^{-c} \right)
\end{equation*}
Hence, we have weak convergence of $S_n$ to $S$ as long as the slowest term converges. The relative rates at which the $n_j$'s grow to infinity are not restricted. 

For comparison, in OLS on units with cluster dependence, observations are not standardized within each cluster. As a result,  the contribution of each cluster to ``numerator" in the $\hat{\beta} - \beta$ is $X_j'U_j$ rather than $\frac{X_j'U_j}{\sqrt{n_j}}$, where $X_j$ is the stacked covariates for units in cluster $j$ and $U_j$ are their stacked linear regression errors. Hence,  large clusters have outsize influence in estimation and inference. Restricting the influence of each cluster motivates the restricted heterogeneity assumptions in \cite{hl2019} for example.

\section{Cluster Statistics for \cite{gllqsx2019}}\label{appendix--clusterstatistics}

\begin{table}[htbp]
      \begin{minipage}{.5\linewidth}
  	\footnotesize 
      \begin{tabular}{cccc}
       \cmidrule{1-4}\morecmidrules\cmidrule{1-4}        
      School & Track & Group & Size \\
      \midrule
      \multirow{27}[6]{*}{US 1} & \multirow{4}[2]{*}{Honors} & 1     & 325 \\
            &       & 7     & 350 \\
            &       & 11    & 625 \\
            &       & 27    & 725 \\
  \cmidrule{2-4}          & \multirow{15}[2]{*}{Regular} & 2     & 300 \\
            &       & 3     & 325 \\
            &       & 4     & 400 \\
            &       & 6     & 250 \\
            &       & 9     & 225 \\
            &       & 10    & 300 \\
            &       & 12    & 250 \\
            &       & 13    & 350 \\
            &       & 14    & 375 \\
            &       & 15    & 500 \\
            &       & 17    & 375 \\
            &       & 22    & 275 \\
            &       & 24    & 300 \\
            &       & 26    & 325 \\
            &       & 28    & 275 \\
  \cmidrule{2-4}          & \multirow{8}[2]{*}{Others} & 5     & 225 \\
            &       & 8     & 250 \\
            &       & 18    & 400 \\
            &       & 19    & 150 \\
            &       & 23    & 450 \\
            &       & 25    & 150 \\
            &       & 29    & 25 \\
            &       & 30    & 25 \\
      \midrule
      \multirow{2}[3]{*}{US 2} & Honors & -     & 46 groups of 25 \\
  \cmidrule{2-4}          & Regular & -     & 60 groups of 25 \\
       \cmidrule{1-4}\morecmidrules\cmidrule{1-4}    
      \end{tabular}%
  	\end{minipage}
    \begin{minipage}{.5\linewidth}
  	\footnotesize  
    \begin{tabular}{cccc}
     \cmidrule{1-4}\morecmidrules\cmidrule{1-4}        
    School & Year  & Group & Size \\
    \midrule
    \multirow{2}[2]{*}{Shanghai 1} & \multirow{2}[2]{*}{2016} & 9992  & 750 \\
          &       & 9993  & 750 \\
    \midrule
    \multirow{3}[4]{*}{Shanghai 2} & \multirow{2}[2]{*}{2016} & 9994  & 1000 \\
          &       & 9995  & 1000 \\
\cmidrule{2-4}          & 2018  & -     & 128 groups of 25 \\
    \midrule
    \multirow{5}[4]{*}{Shanghai 3} & \multirow{4}[2]{*}{2016} & 9996  & 975 \\
          &       & 9997  & 975 \\
          &       & 9998  & 800 \\
          &       & 9999  & 750 \\
\cmidrule{2-4}          & 2018  & -     & 122 groups of 25 \\
    \midrule
    Shanghai 4 & 2018  & -     & 126 groups of 25 \\
     \cmidrule{1-4}\morecmidrules\cmidrule{1-4}    
    \end{tabular}%
    \end{minipage}
    \captionsetup{justification=centering}    
    \caption{Cluster Structure for US and Shanghai Schools}
\end{table}%

\end{document}